\documentclass[]{aastex}

\usepackage{emulateapj5}
\usepackage{onecolfloat}
\usepackage{graphicx}
\usepackage{fancyheadings}
\usepackage{ulem}
\usepackage{rotating}
\usepackage{lscape}

\def\NH{$N$(H I)}

\def\etal{et al.}

\def\lya{Ly$\alpha$ }
\def\smpy{M$_{\odot}{\rm \ yr^{-1}}$}
\def\smpykpc{M$_{\odot}{\rm \ yr^{-1} \ kpc^{-2}}$}
\def\smpympc{M$_{\odot}{\rm \ yr^{-1} \ Mpc^{-3}}$}

\def\Rratio{${{\cal R}(v)}$}

\def\kms{km~s$^{-1}$ }
\def\omgm{$\Omega_{{\rm M}}$}
\def\omgv{$\Omega_{\Lambda} $}

\def\micron{$\mu$m}

\def\gamdnr{${\Gamma_{d}}$}
\def\peq{$P_{eq}$}
\def\pmin{$P_{min}$}
\def\pmax{$P_{max}$}

\def\kapnr{${\kappa}$}
\def\cm2{\, \rm cm^{-2}}

\def\perd{\;\;\; .}
\def\cmma{\;\;\; ,}

\def\rhodot{$\dot{\rho_{*}}$}
\def\rhodotz{$\dot{\rho_{*}}$$(z)$}
\def\rhosz{${\rho_{*}(z)}$}
\def\rhos{${\rho_{*}}$}

\def\ps{$\dot{\psi_{*}}$}
\def\ms{$\dot{M_{*}}$}
\def\psav{$<$$\dot{\psi_{*}}$$>$}
\def\psavz{${<{{\dot{\psi_{*}}}}(z)>}$}
\def\ciis{C II$^{*}$}
\def\nh{$N$(H  I)}
\def\lclos{$l_{c}$}

\def\lcrnr{$l_{cr}$}
\def\jnu{$J_{\nu}$}

\def\junit{ergs cm$^{-2}$ s$^{-1}$ Hz$^{-1}$ sr$^{-1}$}

\begin{document}

\twocolumn[%
\submitted{submitted to the Astrophysical Journal Nov.5,2002}

\title{C II$^{*}$ ABSORPTION IN DAMPED \lya\ SYSTEMS: (II) A NEW WINDOW ON
THE STAR FORMATION HISTORY OF THE UNIVERSE}

\author{ ARTHUR M. WOLFE,\altaffilmark{1}\\ 
Department of Physics and Center for Astrophysics and Space Sciences; \\
University of California, San
Diego; \\
C--0424; La Jolla; CA 92093\\
{\bf awolfe@ucsd.edu}}
 
\author{} 

\author{ ERIC GAWISER\altaffilmark{1,2}\\ 
Department of Physics, and Center for Astrophysics and Space Sciences; \\
University of California, San
Diego; \\
C--0424; La Jolla; CA 92093\\
{\bf egawiser@ucsd.edu}}

\author{and} 

\author{JASON X. PROCHASKA,\altaffilmark{1}\\ 
UCO-Lick Observatory; \\
University of California, Santa Cruz\\
Santa Cruz, CA; 95464\\
{\bf xavier@ucolick.org}}

\begin{abstract} 
Starting from the SFR per unit physical area,  determined
for DLAs using the C II$^{*}$ method, we obtain
the SFR per unit comoving volume
at $z$ $\approx$ 3. Pure warm neutral medium (WNM) solutions
are ruled out since they generate more
bolometric background radiation than
observed, but the CNM-dominated
two-phase solutions are consistent with the backgrounds.
We find the SFR per unit comoving volume
for DLAs agrees with that for 
the Lyman Break Galaxies (LBGs). Though the mass of produced
stars indicated by the SFRs is consistent with the current
densities of known stellar populations, the mass of metals produced
by $z$=2.5
is 30 times larger than detected in absorption in 
DLAs. Of the three possible solutions to this ``missing
metals'' problem, the most likely appears to be that star
formation occurs in compact bulge regions.
We search for evidence of feedback and
find no correlations between the SFR per unit area
and N(H I),
but  possible
correlations between SFR per unit area and low-ion
velocity width and SFR per unit area and metal abundance. 
We show that 
(a) the correlation between
cooling rate and dust-to-gas ratio is positive
evidence for grain photoelectric heating,
(b)
the CMB does not significantly populate
the  C II excited fine-structure states, and
(c) the ratio of CII$^{*}$ to resonance-line
optical depths is a sensitive probe of the multi-phase
structure of the DLA gas.
We address recent arguments that
DLAs are comprised only of WNM gas, and show them to
be inconclusive. Despite the rough agreement between
SFR per unit comoving volume for DLAs and LBGs, current evidence indicates
these are distinct populations.

 \end{abstract}


]
\altaffiltext{1}{Visiting Astronomer, W.M. Keck Telescope.
The Keck Observatory is a joint facility of the University
of California and the California Institute of Technology.}

\altaffiltext{2}{Current address: Yale Astronomy Department,
P. O. Box 208101, New Haven, CT, 06520}

\pagestyle{fancyplain}
\lhead[\fancyplain{}{\thepage}]{\fancyplain{}{Wolfe, Prochaska, \& Gawiser}}
\rhead[\fancyplain{}{STAR FORMATION IN DAMPED {\lya} SYSTEMS}]{\fancyplain{}{\thepage}}
\setlength{\headrulewidth=0pt}
\cfoot{}

\section{INTRODUCTION}

This is the second of two papers describing a new method for obtaining
star formation rates (SFRs) in damped {\lya} systems (DLAs). In Paper
I (Wolfe, Prochaska, \& Gawiser  2003) we showed how measurements
of {\ciis} 1335.7 absorption lines in DLAs allow one to infer the cooling
rate per H atom of the neutral gas. Since we assume steady-state conditions,
this equals the heating rate per H atom, which we use to infer the
SFR per unit area, {\ps}. We do this by assuming gas in DLAs to be heated
by the same mechanism responsible for heating the ISM in the Milky Way,
the grain photoelectric effect (Bakes \& Tielens 1994;
Weingartner \& Draine 2001a).  In that case the heating rate is
proportional to the 
product of the dust-to-gas ratio, {\kapnr}, the photoelectric heating
efficiency, $\epsilon$,  and the mean intensity of FUV radiation,
{\jnu}; the latter
is proportional to {\ps} for sources in
a plane parallel layer. Specifically, in Paper I we modeled
DLAs as  uniform gaseous 
disks with radius, $R$, and scale-height $h$, in which
the sources of FUV radiation were uniformly distributed.
We also showed how {\kapnr} can be deduced from
the [Fe/Si] and [Si/H] abundance ratios (recall [X/Y]$\equiv$log$_{10}$(X/Y)
$-$log$_{10}$(X/Y)$_{\odot}$) for the following assumptions about grain
composition: grains were either carbonaceous
as in the Galaxy (the ``Gal'' model) or Silicates as in the SMC
(the ``SMC'' model). We inferred {\kapnr} by assuming the number of
depleted C or Si atoms per depleted Fe atom to be the same in DLAs
as in the ISM. Furthermore, we considered depletion ratios ranging
from a minimal ``nucleosynthetic ceiling'' in which the 
intrinsic ratio, [Fe/Si]$_{int}$=$-$0.2,
to a maximal depletion ratio, [Fe/Si]$_{int}$=0.0
(see Prochaska \& Wolfe 2002; hereafter referred to as PW02).

In Paper I we solved the transfer equation for {\jnu} and then
calculated the thermal equilibrium of gas subjected to cosmic-ray and
X-ray heating in addition to grain photoelectric heating. The gas was
assumed to cool in the usual way; i.e., by emission of fine-structure,
metastable, and {\lya} lines as well as grain recombination
radiation. We found that gas can reside in two thermally stable
states; a cold neutral medium (CNM) and a warm neutral medium (WNM)
(see Wolfire {\etal} 1995, hereafter W95). Typically, the densities
and temperatures of the CNM and WNM are 10 cm$^{-3}$and 150 K, and 0.2
cm$^{-3}$ and 8000 K respectively. We further assumed the CNM and WNM
to be in pressure equilibrium at pressure
$P_{eq}$=$(P_{min}P_{max})^{1/2}$ where $P_{min}$ and $P_{max}$ are
the minimum and maximum pressures of the pressure versus density
curve.  We considered a CNM model in which the line-of-sight to the
background QSO encounters comparable column densities of gas in the
CNM {\em and} the WNM. We also considered a WNM model in which the
line-of-sight encountered only WNM gas at pressure equal to
$P_{eq}$. Combining the measured heating rates with those predicted at
the thermally stable gas densities, $n_{CNM}$ and $n_{WNM}$, we
obtained unique values for {\ps} for each DLA; one value for the CNM
solution and the other for the WNM solution.  We then averaged {\ps},
for two redshift bins centered at $z$=2.15 and $z$=3.70, to derive the
average SFR per unit physical area, {\psavz}. The WNM models result in
significantly higher SFRs than the CNM models since the measured {\\}
[C II] 158 {\micron} cooling rate per H atom, {\lclos}, is a small
fraction of the total cooling rate in the WNM, whereas {\lclos{ equals}}
the total cooling rate in the CNM (see Paper I).

This paper starts by considering quantities with cosmological
significance.  Specifically, in $\S$ 2 we combine {\psavz} with the
incidence of DLAs per unit absorption distance interval, $d{\cal
N}/dX$ (Bahcall \& Peebles 1969), to derive the SFR per unit comoving
volume, {\rhodotz}.  We then derive the bolometric background
intensity, $I_{EBL}$, for all model combinations. We show that the WNM
models produce more background radiation than observed in every case,
and as a result are ruled out. By contrast the CNM models are
consistent with the observed values of $I_{EBL}$. We compute a
consensus model, which is an average over all the CNM models.  We show
that the resulting {\rhodotz} are comparable to {\rhodotz} inferred
for the Lyman Break Galaxies (Steidel {\etal} 1999; hereafter referred
to as LBGs). In $\S$ 3 we consider implications of these results. We
compute the mass of stars and the mass of metals produced by the star
formation history, {\rhodotz}, of the consensus model. While the mass
of stars is consistent with masses of current stellar populations, the
mass of metals produced by $z$=2.5, is more than 30 times the mass of
metals inferred for DLAs at the same redshift. We discuss possible
solutions to this dilemma including a ``bulge'' model in which star
formation is confined to a compact region located at the center of the
extensive region creating {\ciis} absorption.  We consider independent
evidence for (a) star formation and (b) the deposition of stellar
energy into the absorbing gas, i.e., feedback. At this point the
reader not interested in the physics of interstellar gas can turn to
the final section, $\S$ 6.  Having discussed various implications of
our models we proceed to test their self-consistency in $\S$ 4 where
three tests are carried out.  First, we find a statistically
significant correlation between the [C II] 158 {\micron} cooling rate
per atom, {\lclos}, and {\kapnr}, which is strong evidence in favor of
grain photoelectric heating.  Second, we show that the spontaneous
energy emission rate, {\lclos}, reflects the cooling rate of the gas
instead of the excitation level caused by CMB radiation.  Third, we
examine the ratio of {\ciis} to resonance-line optical depths to look
for evidence of shifts in gas phase.  In $\S$ 5 we discuss arguments
made by other authors against the presence of CNM gas in DLAs. A
careful reassessment of these arguments shows that they do no rule out
the presence of CNM gas in DLAs. A summary and concluding remarks are
given in $\S$ 6.

Unless
stated otherwise 
we adopt an Einstein-deSitter cosmology
in which {\omgm}=1, {\omgv}=0, and $h$=0.5 to 
facilitate comparison with published results.

\section{COSMOLOGICAL QUANTITIES}

We now turn to quantities with cosmological significance. We compute the
SFR per unit comoving volume, {\rhodot}. The redshift dependence
of {\rhodot} implies a star formation history throughout spacetime that gives rise to 
background radiation. 
We calculate  the bolometric intensity of this background radiation 
for
the CNM and WNM models and compare the results with observations. 
We then construct a consensus CNM model for {\rhodotz}, which
is consistent with measurements of the background radiation.

\subsection{The  SFR per Unit Comoving Volume}

The SFR per unit comoving volume for DLAs is given by
\begin{equation}
{{\dot{\rho_{*}}}}(z)={<{{\dot{\psi_{*}}}}(z)>}An_{co}(z)
\label{eq:rhostardot1}
\cmma
\end{equation}

\noindent where {\psavz} is the average SFR per unit
physical area at redshift $z$, and $A$ and $n_{co}$ are the
average physical cross-sectional area and comoving density
of the DLAs. While neither $A$ nor $n_{co}$ has been determined from observations,
their product is known from the incidence of DLAs per unit absorption distance
interval, $d{\cal N}/dX$ (e.g. Storrie-Lombardi \& Wolfe 2000). 
We find
\begin{equation}
d{\cal N}/dX=A_{p}n_{co}(X)
\cmma
\label{eq:dNdXdef}
\end{equation}

\noindent where 
$A_{p}$ is the average projection of $A$ on the plane of the sky, and
$X(z)$, the absorption distance (Bennett {\etal} 2003), is given by
\begin{equation}
{dX \over dz} = 
{\bigl(}cH_{0}^{-1}{\bigr)}{\Biggl[}{{(1+z)^{2}} \over {[(1+z)^{2}(1+{\Omega_{{\rm M}}}z)-z(z+2){\Omega_{\Lambda}}]^{1/2}}}{\Biggr]}
\perd
\label{eq:dXdzdef}
\end{equation}

\noindent As a result
\begin{equation}
{{\dot{\rho_{*}}}}(z)={<{{\dot{\psi_{*}}}}(z)>}(A/A_{p})d{\cal N}/dX
\perd
\label{eq:rhostardot2}
\end{equation}

\noindent

We computed {\rhodot} by assuming the DLAs to be plane-parallel layers;
i.e., $A/A_{p}$=2,
and by choosing an Einstein-deSitter cosmology ({\omgm}=1,$\Omega_{\Lambda}$=0,
$h$=0.5). Although this model is ruled out by observations (e.g. Bahcall {\etal} 1999),
it is the model used
in most published determinations of {\rhodot}
(e.g., Steidel {\etal} 1999; Lanzetta {\etal} 2002), and for this
reason we selected it 
for purposes of comparison. 
We chose {\psavz} and its associated errors from Table 3 in Paper I
and calculated 
$d{\cal N}/dX$ at the mean $z$ of the redshift
bins  from the expression 
\begin{equation}
d{\cal N}/dX=(H_{0}/c){0.055{\times}(1+z)^{0.61}}
\cmma
\label{eq:dNdX2}
\end{equation}

\noindent found by Storrie-Lombardi {\&} Wolfe (2000) 
for the Einstein-deSitter cosmology. 
Errors in {\rhodot} were
computed by propagating errors in {\ps} and in $d{\cal N}/dX$.

The results for the ``Gal'' dust models
are shown in Figure 1 as magenta  data points for minimal
depletion 

\begin{figure}[h]
\centering
\scalebox{0.33}[0.25]{\rotatebox{-90}{\includegraphics{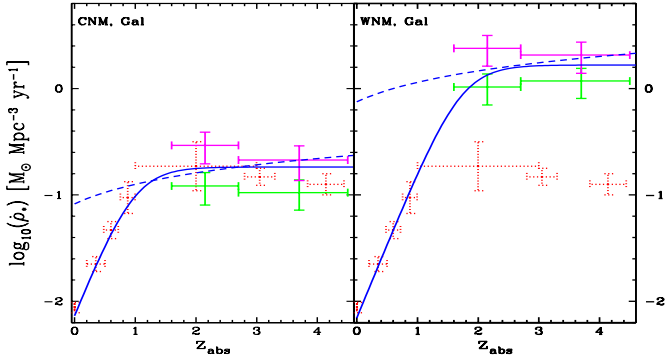}}}
\caption[]{ SFRs per unit comoving volume for ``Gal'' dust model shown
as magenta data points for minimal depletion and green data points for
maximal depletion. Results for CNM model in Figure 1a and for WNM
model in Figure 1b. Red dotted data points depict {\rhodot} inferred
from galaxy luminosities (cf. Steidel {\etal} 1999; Lilly {\etal}
1996; and Barger {\etal} 2000). Solid blue curves are fits to high-z
DLA data that also agree with galaxy data at $z$ $<$1. Dashed blue
curves are fits to medians of DLA SFRs and are extrapolated to low $z$
assuming {\psavz} is constant and combining equations
({\ref{eq:dNdX2}}) and ({\ref{eq:rhostardot2}}).}
\label{rhovszGal}
\end{figure}

\noindent and green data points for maximal depletion.  As expected,
the SFRs per unit comoving volume for the WNM models (Figure 1b) are
at least 10 times higher than for the corresponding CNM models (Figure
1a). Furthermore, for every model, minimal depletion gives rise to
higher {\rhodot} than the maximal depletion.  This is because for a
given heating rate, {\ps} is inversely proportional to the dust-to-gas
ratio, {\kapnr}, and {\kapnr} is lower for minimal depletion than for
maximal depletion. Figure 1 also reveals no evidence for statistically
significant redshift evolution of {\rhodot} determined by the {\ciis}
technique.  This is in accord with determinations of {\rhodot} from
luminosities measured for flux-limited samples of galaxies, shown as
red data points.  For the galaxy sample the SFRs in the two highest
redshift bins are based on Lyman-break galaxies that are luminous at
rest-frame UV wavelengths (Steidel {\etal} 1999), while the four
lowest redshift points are based on galaxies with lower luminosities
(Lilly {\etal} 1996). The bin at $z$ = 2 is based on far-infrared
(FIR) luminous galaxies detected by SCUBA (Holland {\etal} 1999). The
redshifts for these objects were determined from a still uncertain
radio-FIR photometric redshift indicator (Barger {\etal} 2000).
Interestingly, the magnitude of the comoving SFRs deduced by the
{\ciis} and galaxy luminosity techniques are not very different in the
redshift interval where they overlap; i.e., $z$$\approx$[2,4.5].  For
the CNM model the difference is less than a factor of 2, while for the
WNM model the difference is about a factor of 10.  We discuss possible
implications of this agreement in $\S$ 2.3 and $\S$ 6.

To test the generality of these conclusions, the calculations were
repeated for the ``SMC'' dust models. The results, shown in Figure 2,
reveal the same patterns as  
found for ``Gal'' dust. The principal difference
is the systematically higher values of {\rhodot}
predicted by corresponding ``SMC'' models. This is because the
photoelectric
heating efficiency of silicate grains is lower than that of carbonaceous
grains (see Figure 15 in Weingartner \& Draine

\begin{figure}[h]
\centering
\scalebox{0.33}[0.25]{\rotatebox{-90}{\includegraphics{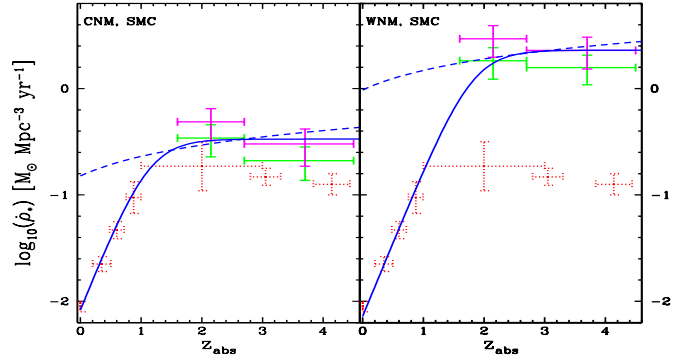}}}
\caption[]{ Same as Figure 1, except ``SMC '' dust model
is assumed.}
\label{rhovszSMC}
\end{figure}

\noindent 2001a), and as a result
higher SFRs are required to achieve a given heating rate in
the case of silicate (i.e., ``SMC'') 
\noindent grains. Comparison between Figures
1 and 2 also reveals smaller differences between {\rhodot}
derived for minimal and maximal depletion in the case of ``SMC'' dust
than for ``Gal'' dust. The phenomenon is present in the CNM
model but not the WNM model. In the case of maximal depletion,
[C/H]$_{gas}$ is larger by 0.2 dex than for minimal depletion.
But an increase in [C/H]$_{gas}$  causes a decrease
in $n_{CNM}$, which 
can increase {\ps} significantly if $n_{CNM}$
intersects the rising {\lcrnr}($n$) curves caused by the transition
from WNM to CNM temperatures (see Figure 5 in Paper I). In the case
of ``SMC'' dust
this effect compensates for the decrease in {\ps}
caused by the increase in {\kapnr} discussed above,
because the heating rate {$\Gamma_{d}$} 
increases more rapidly with density than for ``Gal'' dust 
(Weingartner \& Draine 2001a).  
Consequently, the net difference in {\ps}, hence {\rhodot}, is smaller for ``SMC''
than for ``Gal'' dust.



\subsection{Bolometric Background Intensity}

While
the {\em general} trends in the {\rhodot} versus $z$ plane
appear to be insensitive to our choice of model assumptions,
large systematic uncertainties in the comoving SFRs remain.
Though uncertainties in dust composition and depletion level are
contributing factors, the largest source of error stems from uncertainties
in the thermal phase: Does {\ciis} absorption arise from gas in the CNM or WNM phase?
To address this question we compute the extragalactic bolometric
background intensity, $I_{EBL}$. Because $I_{EBL}$ is generated
by a given star formation history, and since the star formation
histories indicated by the CNM and WNM models are very different,
measurements of the background intensity may be able to discriminate
between them.

The bolometric extragalactic background intensity generated by a given
star
formation history, {\rhodot}($z$), is as
follows:
\begin{equation}
I_{EBL}={c \over 4{\pi}}{\int_{t_{F}}^{t_{0}}{dt \over 1+z(t)}}\int_{0}^{t}{{\dot{\rho_{*}}}}(t-t^{\prime})L(t^{\prime})d{t^{\prime}} 
\cmma
\label{eq:IEBL}
\end{equation}

\noindent where $L({t^{\prime}})$ is the bolometric luminosity per unit mass as a function
of age {$t^{\prime}$} of a stellar population with a specified
IMF (Madau \& Pozzetti 2000),
$t_{F}$ is the formation epoch of the stellar population, and $t_{0}$ 
is the current age of the universe. We tested the models by
computing backgrounds generated by  
analytic fits to the {\ciis} comoving SFRs. The fits are
shown as smooth curves 
in Figures 1 and 2, and the backgrounds they generate
as corresponding curves in Figure 3.

First, we explored the hypothesis that high-$z$ DLAs evolve into
normal low-$z$ galaxies. Evolution into normal galaxies is 
consistent with the recent identification of DLAs at $z$ $<$ 1
with the local galaxy population (Zwaan {\etal} 2002; Turnshek
{\etal} 2002; Rosenberg \& Schneider 2003). 
In that case the star formation histories are 
constrained to pass through the comoving SFRs measured for
galaxies at $z$ $<$ 1 and  for DLAs at higher redshifts.
The results are shown as solid blue curves for ``Gal'' dust
in Figure 1 and ``SMC'' dust in Figure 2. For the CNM model
(Figures 1a and 2a) the curves resemble the star formation
histories inferred from galaxy luminosities in the
redshift interval  $z$ = [0,5].
For the WNM model (Figures 1b and 2b) the fits greatly exceed
the comoving SFRs inferred for the LBGs at
$z$ $>$ 2, though they are in
good agreement with the galaxy data at $z$ $<$ 1.
Here the DLAs could represent a population of objects 
undetected in emission at high $z$ that evolve into normal galaxies
at low redshifts. Second, we considered an hypothesis
in which the star formation histories of DLAs are dictated solely
by their redshift evolution, without regard to the
galaxy data. In this case we
combined the expression for {\rhodot}($z$) 
in equation ({\ref{eq:rhostardot2}}) with the expression for
$d{\cal N}/dX$ in equation ({\ref{eq:dNdX2}}). Although
we find $<${\ps}$(z)>$ is independent  
of redshift in the redshift interval $z$ = [1.6,4.5], at lower
redshifts there are no
measurements of
{\ciis}, and as a result
$<${\ps}($z$)$>$ is unknown.  
Low redshifts are crucial
because that is where most of the background radiation arises.
For simplicity we let $<${\ps}$(z)>$ equal a  constant 
evaluated by averaging over all the {\ps} inferred
from {\ciis} absorption. 
The results are shown as dashed curves in Figures 1 and 2.

The resulting backgrounds are shown in Figure 3 where we plot the
predicted bolometric intensity from DLAs with $z$ $\ge$ $z_{min}$
versus $z_{min}$.  The significance of this quantity,
$I_{EBL}(z{\ge}z_{min})$, is that it reveals the contribution to the
measured background, $I_{EBL}$ (i.e, $I_{EBL}(z{\ge}0)$), made by DLAs
in the redshift range for which {\rhodot}($z$) has been determined;
i.e., $z$=[1.6,4.5].  The backgrounds were obtained with an
Einstein-deSitter cosmology assuming $h$ = 0.5, by adopting the same
IMF used to relate {\ps} to mean intensity (see Paper I), using
Bruzual and Charlot's (1993) population synthesis libraries (see Madau
\& Pozzetti 2000), and by assuming a formation redshift, $z_{F}$ = 5.
For comparison, the two horizontal straight lines depict upper and
lower limits on $I_{EBL}$ set by measurements between 0.15 {\micron}
and 1000 {\micron} (Hauser \& Dwek 2001). This wavelength range is
relevant since it brackets the background spectra predicted for most
models of DLAs (e.g. Pei {\etal} 1999).  According to Dwek (2002) the
upper limits, which are crucial here, are conservative and should be
regarded as 95 $\%$ confidence limits.  It is important to emphasize
that {\em while {\rhodotz} is cosmology dependent, the backgrounds
computed from the {\ciis} technique (i.e. by combining equations
(\ref{eq:rhostardot2}) and (\ref{eq:IEBL})) are independent of the
adopted cosmology and Hubble constant. Therefore, consistency between
theory and observation is independent of the cosmology one assumes.}

The background intensities predicted for the WNM models are too high.
Figures 3b and 3c depict backgrounds generated by the ``Gal'' and
``SMC'' dust models

\begin{figure}[h]
\centering
\scalebox{0.35}[0.40]{\rotatebox{-90}{\includegraphics{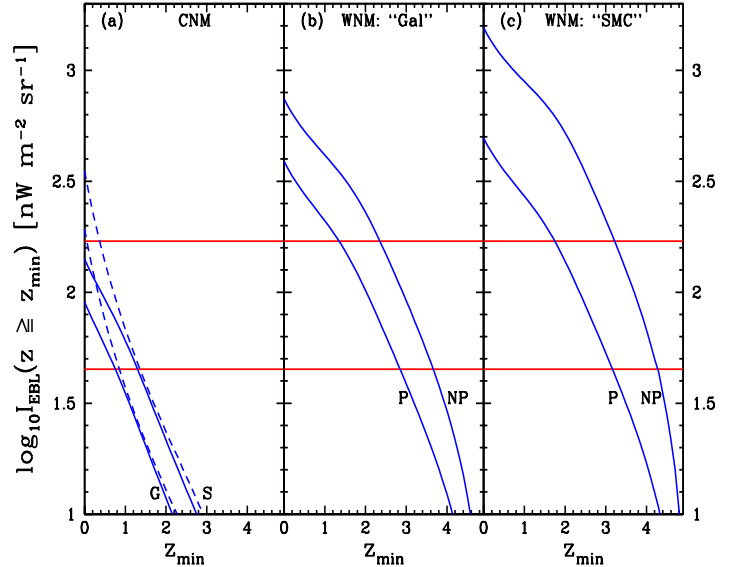}}}
\caption[]{ Smooth curves depict bolometric background intensity due to
DLAs with $z$ $\ge$ $z_{min}$. Horizontal lines are empirical 95 $\%$
confidence upper and lower limits. Figure 3a shows solutions for CNM
model. Dashed curves labeled ``G'' and ``S'' are generated by ``Gal''
and ``SMC'' star formation histories shown as dashed curves
in Figures 1a and 2a.
Solid curves so labeled are generated by ``Gal'' and ``SMC'' star
formation histories shown as solid curves in Figure 1a and 2a. Figure
3b shows solutions for WNM model for ``Gal'' dust. Curve labeled ``P''
is background generated by solid curve in Figure 1b, in which
optical pumping is included. Curve labeled ``NP'' is result
withoug optical pumping. Figure 3c is the same as 3b, except
that the ``SMC'' star formation history of Figure 2b is used.}
\label{rhovszGal}
\end{figure}

\noindent respectively. The curves labeled ``P'' correspond
to the star formation histories generated by the solid curves in
Figures 1b and 2b, where ``P'' indicates that optical pumping is
included (where optical pumping is the mechanism by which
the populations of the ground-term fine-structure states are mixed
through UV excitations to higher lying levels [e.g. Sarazin {\etal}
1979]).  The curves labeled ``NP'' show backgrounds generated by
star-formation histories without optical pumping.  As discussed in in
Paper I the true WNM solution lies between these limits. The predicted
background for the ``Gal'' model is between 2.5 and 5 times the 95
$\%$ confidence upper limit on $I_{EBL}$, and between 3 and 10 times
this limit for the ``SMC'' model. In both models, sources with
$z$$\ge$ $z_{min}$=1.6 generate background intensities exceeding the
95 $\%$ confidence upper limit, indicating that DLAs in which {\ciis}
absorption arises in WNM gas produce more background radiation than
observed. When the contribution of lower-$z$ galaxies is included, the
observed upper limits to the backgrounds are exceeded by much larger
factors.  Had we computed backgrounds generated by the dashed curves
in Figures 1b and 2b, we would reach the same conclusion; i.e., the
WNM models are ruled out.


This disproof of the WNM-dominated solution appears to be robust.  The
FUV mean intensities inferred for 4 DLAs in which ${\rm H}_{2}$ is
detected (Ge \& Bechtold 1997; Srianand {\etal} 2000; Molaro {\etal}
2002; Levshakov {\etal} 2002) are comparable to $G_{0}$=1.7, the value
found for the Galaxy ISM (Draine 1978; note $G_{0}$ is {\jnu} in
convenient units of 10$^{-19}$ ergs cm$^{-2}$ s$^{-1}$ Hz$^{-1}$
sr$^{-1}$).  Although these results may need to be corrected for
suppression of optical pumping (Sarazin {\etal} 1979), the implication
is that {\psav} is significantly lower than required by the WNM
models.  Furthermore, the backgrounds predicted for the WNM models are
conservative lower limits. This is because we assumed the pressure of
the two-phase medium, $P_{eq}$, to exceed $P_{min}$, the maximum
pressure allowed for gas in a pure WNM phase.  Values of $P_{eq}$ $<$
$P_{min}$ would result in lower values of $n_{WNM}$, hence higher
{\rhodot}. Note, that high values of {\rhodot} are predicted even if
{\ciis} absorption arises in warm gas with temperatures below that
predicted by our two-phase model. We considered scenarios in which
{\ciis} absorption arises in gas with $T$$\sim$1000 K; i.e., in
thermally unstable gas like that predicted by Vazquez-Semadeni {\etal}
(2000).  Figure 3c in Paper I shows that at such temperatures the
total cooling rate is considerably larger than the 158 {\micron}
emission rate, especially in the absence of optical pumping. The
resulting backgrounds are significantly above the 95 $\%$ confidence
upper limit on $I_{EBL}$ when the emission from $z$ $<$ 1 galaxies is
added to the contribution from DLAs.  Therefore, WNM models, or any
model in which [C II] 158 {\micron} emission does not dominate the
cooling rate, are unlikely to be correct.

By contrast, CNM models in which {\psavz} decreases with decreasing
redshift are consistent with the background data. Consider models in
which {\psav} equals a constant given by the average of all the
inferred values of {\ps} for the CNM models.  These are depicted by
dashed curves in Figure 3a and correspond to the ``Gal'' ($G$) and
``SMC'' ($S$) star formation histories shown as dashed curves in
Figures 1a and 2a. Both models predict $I_{EBL}(z{\ge}0)$ to be above
the 95 $\%$ confidence upper limit on bolometric intensity. Moreover,
Figures 1a and 2a show that {\rhodot} at $z$=0 is significantly above
the comoving SFR inferred from the luminosities of local galaxies.  On
the other hand, the solid curves in Figure 1a and 2a depict model star
formation histories that are compatible with {\rhodotz} inferred for
galaxies at $z$ $<$ 1.  These curves generate the backgrounds shown as
solid curves in Figure 3a, which are compatible with the limits on
$I_{EBL}$.  For these models {\psavz} at $z$ = 0 must be significantly
below the $\approx$ 10$^{-2.2}$ {\smpykpc} determined at high
redshifts both for ``Gal'' and ``SMC'' dust. At first, this appears to
be inconsistent with the observation that {\psav} deduced for local
disk galaxies is comparable to 10$^{-2.4}$ {\smpykpc} (Kennicutt
1998).  But the latter star formation estimates were made by averaging
over the corrected de'Vaucouleurs' radius, $R_{0}$, whereas DLA
absorption at any redshift would occur out to an average radius, $R_{H
I}$$\approx$2$R_{0}$ (Wolfe {\etal} 1986).  However, to compute
{\rhodot}, it is necessary to average {\ps} over $R_{H I}$, not
$R_{0}$.  Therefore, the appropriate value of {\psav} for local disk
galaxies should be reduced by a factor of four to 10$^{-3.0}$
{\smpykpc}.  {\em We conclude that if DLAs evolve into normal
galaxies, their SFR per unit area has decreased significantly since
$z$ $\approx$ 1.6. We emphasize this conclusion holds for {\ps}
averaged over $R_{H I}$ rather than $R_{0}$, which is normally used
for computing {\ps} in nearby galaxies (Kennicutt 1998)}.
\footnote{When we compared the average {\lclos} in DLAs to
the average {\lclos} for the ISM (Paper I), we found the
ratio, ({\lclos})$_{DLA}$/({\lclos})$_{ISM}$ roughly equaled
the average dust-to-gas ratio of DLAs relative to that of 
the ISM, i.e. {\kapnr}. This indicated $G_{0}$ in DLAs
to be roughly equal to the ISM value, $G_{0}$=1.7, which
corresponds to {\ps}=10$^{-2.4}$ {\smpykpc}. Does this
contradict our finding that {\ps} at $z$=0 equals
10$^{-3.0}$ {\smpykpc}? The answer is no because the high
opacity of dust in the ISM indicates $G_{0}$ arises mainly
from local sources within the optical radius, whereas
the lower opacity of dust in DLAs indicates $G_{0}$
inferred from {\ciis} absorption lines is a global 
average, which  includes the very low SFRs occurring 
outside the optical radius}

\subsection{Consensus Star Formation Model}

By ruling out the WNM hypothesis, we have eliminated half the models
discussed so far. Nevertheless, 
the remaining CNM models contain significant systematic uncertainties, as indicated by
the scatter amongst 
{\rhodotz} inferred from the various dust hypotheses (Figures 1a, 2a).
Here we attempt to assess 
these errors as well as  errors due to other
effects, and to deduce
consensus values for {\rhodotz}.

To estimate the size of the systematic errors, we test the sensitivity
of the CNM models to variations of crucial input parameters.  We find
{\rhodotz} to be sensitive to changes in equilibrium pressure, {\peq},
and that the effect is similar in magnitude to the scatter in
{\rhodotz} due to uncertainties in the composition and depletion level
of dust.  The results in Figures 1 and 2 were computed assuming
$P_{eq}$ = $(P_{min}P_{max})^{1/2}$. Because of the uncertainties
surrounding this criterion (see discussion in Paper I), we now
consider the effects of letting $P_{eq}$ vary between $P_{min}$ and
$P_{max}$.  We find that {\rhodot} decreases with increasing
{\peq}. As {\peq} rises, $n_{CNM}$ increases, which results in lower
values of {\ps} for a fixed {\lclos} (see Figure 5 in Paper I).
Therefore, {\rhodot} is a minimum when {\peq}=$P_{max}$, and a maximum
when {\peq} =$P_{min}$. We assume the variances in {\rhodot} are
determined by differences between these limiting values of {\rhodot}
and the means defined by {\peq}=$(P_{min}P_{max})^{1/2}$.  It is
possible to increase the variances in {\rhodot} further by relaxing
the standard ratio of cosmic-ray ionization rate, $\zeta_{CR}$, to the
SFR per unit area, {\ps}, given in equation (9) in Paper I.  But the
consequent increase in {\rhodot} is constrained by the upper limit on
$I_{EBL}$ to log$_{10}${\rhodot} $<$ $-$0.2 {\smpympc}, while the
decrease is limited to log$_{10}${\rhodot} $>$ $-$1.5 {\smpympc} by
the observed ratios of C II to C I column densities, which become too
small when $\zeta_{CR}$/{\ps} is more than twice the standard value
(see $\S$ 5.1).  Though it is possible for {\rhodot} to attain these
extreme values, it is more likely to remain within the standard
limits, which justifies their use in constructing the
variances. Similar procedures were used to compute variances in
{\rhodot} due to variations in other input parameters such as the
ratio of the radius to height, $R/h$, of the model uniform disk (see
Figure 4 in Paper I).

	In order to compute consensus values of {\rhodotz} in a given
redshift bin, we considered all four possible combinations of ``Gal''
versus ``SMC'' dust composition and minimal versus maximal depletion.
For each of these four models we determined the probability
distribution of {\rhodotz} using the best-fit value and an error
budget that included (a) varying {\peq} from {\pmin} to {\pmax}, (b)
varying the aspect ratio $R/h$ between minimum and maximum values, and
(c) the random errors appearing in Figures 1 and 2.  Although we
suspect that ``SMC'' dust and minimal depletion are more likely to be
correct, we conservatively assumed that all four models were equally
likely. We performed a Monte Carlo simulation drawing an equal number
of simulated

\begin{figure}[h]
\centering
\scalebox{0.35}[0.39]{\rotatebox{-90}{\includegraphics{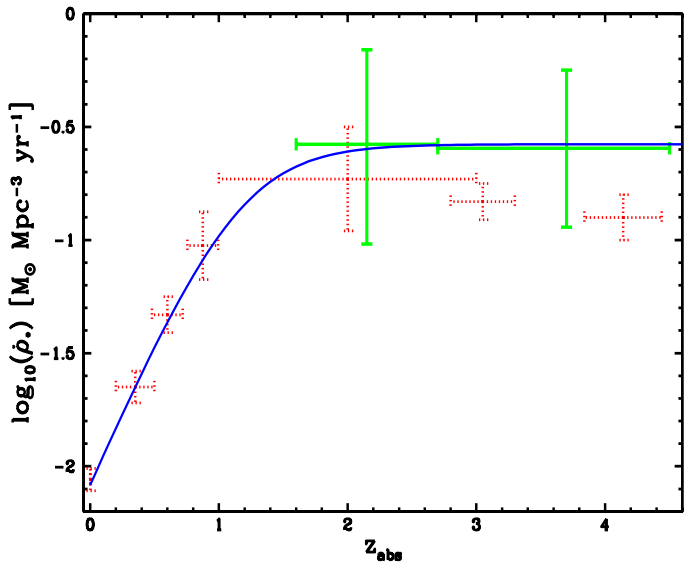}}}
\caption[]{ Green data points depict {\rhodot} and 68 $\%$
confidence errors for ``consensus'' model described in $\S$ 6.3.
Dotted data points are galaxy data described in previous figures. 
Smooth curve is eyeball fit to ``consensus'' model at high
$z$ and galaxy data at low $z$.}
\label{rhovszSMC}
\end{figure}

\noindent data points from each of the four model probability
distributions.  Note that this is equivalent to a Monte
Carlo simulation where each DLA is analyzed using all four models and
then these 4$\times$$n_{DLA}$ data points are resampled at random to
generate the maximum possible variance.  The resulting probability
distribution for {\rhodotz} is well described by a Gaussian, and we
computed the resulting mean, 68\% confidence intervals, and 95\%
confidence intervals.  There are additional systematic uncertainties
that we are unable to quantify at present, including those due to
uncertainties in the grain size distribution, and others that we are
unable to compute such as possible radial variations in {\ps}.  We do
not expect these additional sources of error to dominate.  In
particular, we show in $\S$ 3.2 that the error due to our assumption
of spatially uniform {\ps} is probably less than 0.1 dex.  The largest
systematic uncertainty at present is produced primarily by the model
with ``Gal'' dust composition and maximal depletion, so falsifying
either the ``Gal'' or maximal depletion solutions would raise the
result for {\rhodotz} and significantly reduce the uncertainty.

\begin{table} 
\begin{center}
\begin{tabular}{lcc}
&\multicolumn{2}{c}{log$_{10}${\rhodot}}   \\
\cline{2-3}
&\multicolumn{2}{c}{M$_{\odot}$yr$^{-1}$Mpc$^{-3}$}   \\
\cline{2-3}
$z$&\multicolumn{1}{c}{Einstein deSitter$^{a}$}&\multicolumn{1}{c}{Standard $\Lambda$$^{b}$}  \\
\tableline
2.15 & $-$0.58$\pm$0.42&$-$0.68$\pm$0.42  \\
3.70 & $-$0.59$\pm$0.35&$-$0.71$\pm$0.35  \\
\end{tabular}
\end{center}
\caption{Cosmology Dependence of {\rhodot}} \label{Cosmo}
\tablenotetext{a}{Cosmology with $\Omega_{m}$=1, $\Omega_{\Lambda}$=0, $h$=0.5}
\tablenotetext{b}{Cosmology with $\Omega_{m}$=0.3, $\Omega_{\Lambda}$=0.7, $h$=0.7}
\end{table}

The consensus ``Madau'' diagram for DLAs is shown in Figure 4. The
results are for the Einstein deSitter cosmology, and the interested
reader is referred to Table 1 where we compare these with {\rhodot}
inferred for the standard $\Lambda$ cosmology; the differences are of
order 0.1 dex.  The error bars (corresponding to 68 $\%$ confidence
levels) are of course larger than in Figures 1 and 2, which include
only random errors. Though our errors are larger than reported for the
Lyman-Break galaxies, the latter errors do {\em not} include
systematic errors such as extinction corrections to galaxy
luminosities, which are surely present (see discussion in Steidel
{\etal} 1999). By contrast the effects of dust are essential features
of our models.  Moreover the low values of {\kapnr} imply that at
least half of the radiation from the disk is emitted at rest-frame FUV
wavelengths; i.e., our dust correction is less than a factor of 2.
The blue curve is our eyeball fit through the DLA and low-$z$ galaxy
data in this diagram and will be used in the following section to
compute integrated quantities such as the mass in stars and metals
produced over various time scales.  Of course, it is possible that
high-$z$ DLAs do {\em not} evolve into low-$z$ DLAs and their
associated galaxies, but instead evolve into a population of objects
with luminosity density far below that of normal galaxies. While we
cannot rule out this scenario altogether, we believe it is
implausible. The principal argument against it is the agreement
between the comoving mass density of neutral gas in high-$z$ DLAs and
the mass density of visible matter in current galaxies
(Storrie-Lombardi \& Wolfe 2000). This indicates a connection between
DLAs at high redshift with those at low redshift, unless one assumes
this agreement is a random coincidence.  As a result, the most likely
scenario is one in which {\psavz} decreases in time at $z$ $<$ 1.6 in
such a way that the DLAs evolve into low-$z$ galaxies.

Figure 4 also shows consistency with
{\rhodotz} inferred from gas consumption in DLAs (Pei {\etal} 1999),
which lends credibility to the idea that the decline with time of 
the comoving density of neutral gas is related to star formation.
Furthermore,
Figure 4 indicates approximate agreement between {\rhodotz} determined
for DLAs and LBGs.
{\em That  measurements
of the same quantity by independent techniques based on different
physical considerations are
even in approximate agreement
is either a coincidence or
indicates a connection between DLAs and LBGs.} We shall address this
issue in $\S$ 6.

\section{IMPLICATIONS}

In this section we discuss several consequences of 
this work. In particular we discuss (a) the production of stars
and metals implied by the derived {\rhodotz}, (b) a scenario
in which star formation is confined to a centrally located bulge,
and (c) evidence for feedback.

\subsection{Baryon and Metal Production}

Having determined {\rhodotz} for DLAs, we can
integrate under the smooth curve in Figure 4 to
obtain {\rhosz}, the comoving mass density of
stars at redshift $z$. We find
\begin{equation}
{\rho_{*}(z)}=\int_{z}^{z_{F}}{{\dot{\rho_{*}}}}|dt/dz|dz 
\cmma
\label{eq:rhostar}
\end{equation}

\noindent where this expression for {\rhosz} 
is independent of cosmology and Hubble constant
when {\rhodotz} is determined from the {\ciis} technique; i.e., from
equation ({\ref{eq:rhostardot2}}). Note that unlike $I_{EBL}$, this integral
receives considerable weight from 

\begin{figure}[ht]
\includegraphics[height=3.5in, width=2.9in]{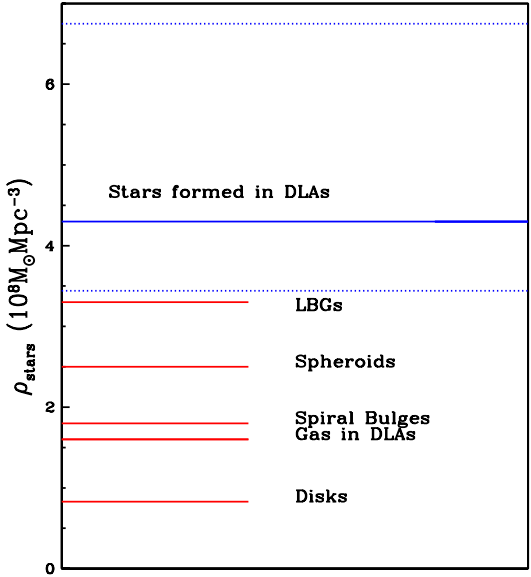}
\caption{Mass density of stars per unit comoving volume
for various populations. Solid blue horizontal line depicts
density produced by SFR for DLAs. Dotted blue lines
are 68 $\%$ confidence level error bars. Red horizontal
lines depict corresponding densities for stars
formed by LBGs,
current spheroids,
current spiral bulges, and current spiral disks.
``Gas in DLAs'' corresponds to comoving mass density of gas in DLAs
at $z$ $\approx$ 3.}
\label{metals density}
\end{figure}

\noindent $z$ $>$ 1.6. 
Pettini (1999) used the last equation to determine
{\rhos}(0), the current mass density in stars formed by the LBGs.
Our estimate of {\rhos}(0) for DLAs,
which we obtain by integrating the last
equation from $z$=0 to $z_{F}$, is shown in Figure 5 along with
68 $\%$ confidence intervals. We compare this to 
determinations of the current mass densities of stars formed by
LBGs and other stellar populations (Fukugita {\etal} 1998).
In deriving this result we (a) integrated to the present under the solid curve
shown in the figure, and (b)
multiplied by Leitherer's (1998) correction
factor of 0.4 (adopted to correct for a more realistic IMF  
[see Pettini 1999]).
Figure 5 shows DLAs
and LBGs
produce the same mass in stars to within 1 $\sigma$.
Moreover, the star formation history of
DLAs suffices to produce the observed stellar content
of spheroids, bulges of spirals, and spiral disks.  
Though the indicated uncertainties
in {$\rho_{*}$(0)}  are large,
the similarity between the predicted stellar content of DLAs
and observed stellar content of galaxies is consistent with the idea that their progenitors
were DLAs
(e.g. Wolfe 1995). In addition, the similarity between the comoving gas density
in DLAs at $z$ $\sim$ 3 (Storrie-Lombardi \& Wolfe 2000) 
and {$\rho_{*}$(0)} is further evidence of self-consistency, though
some infall might be required if the gas density is really lower than {$\rho_{*}$(0)}.	

We also updated Pettini's (1999) calculation for the mass of
metals produced by $z$ = 2.5, the median redshift of DLAs for which
metal abundances have been determined (Pettini {\etal} 1994; PW02).
Pettini (1999) obtained this result using comoving SFRs for 
LBGs whereas we use comoving SFRs for DLAs.
The result shown in Figure 6
was computed assuming
{${\dot{\rho}}_{metals}$}
=(1/42){\rhodot} (Madau {\etal} 1996),
and is compared with the comoving density of

\begin{figure}[ht]
\includegraphics[height=3.5in, width=2.9in]{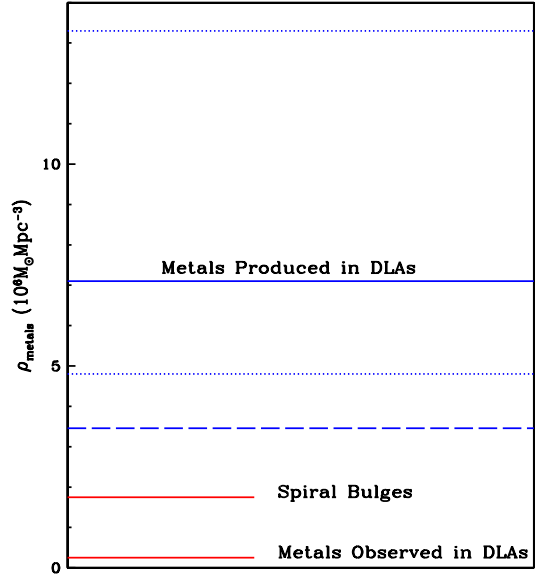}
\caption{Comoving mass density of metals.
Solid blue line shows metals produced in
DLAs by $z$ = 2.5. Dotted blue lines are corresponding 68 $\%$
confidence contours, and dashed blue line is lower 95 $\%$
contour. Red lines correspond to density of metals
in spiral bulges and
in DLAs.}
\label{star density}
\label{metals density}
\end{figure}

\noindent  metals in $z$ = 2.5
DLAs and with the current mass density of metals     
in spiral bulges.
Clearly the metals produced are sufficient to account for the metal
content of spiral bulges. However, as discussed by Pettini (1999)
the mass density of produced metals is 30 times
higher than metals observed
in DLAs. The difference is significant, 
since as shown in Figure 6, the observed metal
content of DLAs is well below the 95$\%$ confidence
contour predicted for the produced metal content.
Therefore, the difference between observed and produced 
metal content is real and leads to a ``missing metals'' problem.
Pettini (1999) first noticed this problem when he  
found that the metals produced
in LBGs exceeded the metals measured in DLAs. 
The problem is much more severe in 
our case because
we are measuring both metal production and metal content in the
{\em same population}.

Three possible solutions to the missing metals problem 
are to sequester 
the metals produced away from the DLA gas observed at $z \sim 2.5$, 
either by confining these metals in ``bulges'', in different systems,
or in the IGM.  
The first solution requires that 
most of the star formation we are seeing is occurring in a compact 
region, i.e., a bulge, and metals produced in this region do not 
rapidly enrich the rest 
of the gas of the galaxy beyond the low metallicities observed 
in DLAs.  The second solution requires that DLAs are a transitory 
phase early in the formation of galaxies, meaning that by the time 
significant metals have been produced the neutral gas has already 
been used up.  Hence the objects observed as DLAs at $z \sim 2.5$ 
are entirely distinct from objects that follow the DLA star 
formation history starting at formation redshifts $z > 4$. 
The third solution solution is to
allow the metal-enriched
material of supernovae to blow out of the DLA galaxy (Mac Low \& Ferrara 1999).
We see two problems with blow out. First, the efficiency of ejecting
metal-enriched gas must exceed 1$-$(29/30)=0.97, which is much larger
than the maximum efficiency of 0.5 seen in local starburst galaxies
(Martin 2003). Second, such ejection would result in a mean
IGM metallicity, [M/H]=$-$1.2, which is at least two orders of magnitude
larger than the metallicity of the {\lya} forest
(Songaila 2001); this would require placing most of the metals
in the high$-$z IGM in some undetectable state.
In the next subsection we shall explore  the solution
we find most appealing,
the bulge hypothesis. 
Other possible explanations 
include changing the IMF from that assumed by Pettini (1999) (which is
the same Madau {\etal} [1996] IMF used in our previous computations)  so
that lower masses of metals are released at the endpoints of stellar evolution.

\subsection{Bulge Hypothesis}

Suppose
star formation is concentrated in the central regions of DLAs, e.g., 
in proto-bulges, which at early times could be configurations of 
molecular clouds that are rarely detected in DLAs because of a small
covering factor or possible obscuration by dust.
In that case most of the metals would be released in the bulge,
which would explain why the mass of metals
produced in DLAs is consistent with the mass of metals in spiral bulges
within the errors, as shown in Figure 6.
The Milky Way bulge is relevant since it
is a metal-rich but old population of stars most of 
which formed by $z$ = 2.5 (Wyse {\etal} 1997). In this
picture 
a small fraction of the metals produced would find their way to the
outer disk via stellar winds or supernovae explosions, thereby explaining
the lower metallicities of the DLA gas.
This idea is self-consistent since, as we now show, the higher
star formation rate per unit area in the central region can account for the
heating rates inferred from the {\ciis} observations of the outer disk
without significantly increasing {\rhodot}.

For simplicity, let the bulge be a sphere with
radius $R_{B}$, which is located at the center of 
a uniform disk with radius $R$ and half-thickness $h$.
In this scenario
the disk
gives rise to damped {\lya} and 
{\ciis} absorption, while FUV radiation  {\em emitted} by the bulge
is the source of the mean intensity at radius $r$, $J_{\nu}^{B}(r)$,
 which heats
the gas through grain photoelectric emission.
In that case
the mean intensity is given by
\begin{equation}
J_{\nu}^{B}(r)={L_{\nu} \over (4{\pi})^{2}r^{2}}{\rm exp}(-k_{\nu}{r})
\cmma
\label{eq:JnuB}
\end{equation}

\noindent where $R_{B}$ $<$ $r$ $<$ $R$,
 $L_{\nu}$ is the Luminosity per unit frequency bandwidth of FUV
starlight emitted by the bulge, 
and $k_{\nu}$ is the absorption opacity due to
dust in the uniform disk. 

The radiation intensity inferred from DLAs by our method 
represents an average over all possible 
lines of sight through these uniform disks, where the C II* column 
determined for each system represents the average of all gas along 
that particular line of sight.  We approximate this ``average 
of averages'' as a simple average of the radiation intensity received at all 
points in the disk, i.e. 

\begin{eqnarray}
< J^{B}_\nu >  
= {\rm C} \int_0^{2 \pi} d\phi {\Biggl [}  
\int_0^{\theta_c} \int_{R_B}^{h \sec \theta} r^2 \sin \theta d \theta dr 
L_\nu g(r) + \nonumber\\
\int_{\theta_c}^{\pi/2} \int_{R_B}^{R \csc \theta} r^2 \sin \theta d \theta dr 
L_\nu g(r) 
{\Biggr ]} 
\label{eq:JnuB_average}
\end{eqnarray}

\noindent with $C=2/[{\pi}R^{2}{2h}{4{\pi}}]$,
$g(r)={\rm exp}(-k_{\nu}r)/(4{\pi}r^{2})$, 
and where we have assumed $R_B<h$, which seems reasonable for the thick 
disks expected at high redshift.
Except for the $R_B$ limit, this 
expression is equivalent to the solution
for $J_{\nu}$ obtained for uniform disks 
(see eq. 14 in Paper I) since 
the luminosity density of 
such disks, 
$\rho_{\nu}$=$L_{\nu}$/($\pi$$R^{2}$2$h$). 
This occurs because
of the symmetry 
between the flux received at the center of a uniform disk from all 
points within the disk and the average flux received over a uniform disk 
from a bulge located at its center.  When integrated,
the solution looks just like that 
of equation 15 in Paper I except that the
1 is replaced by $\exp(-k_{\nu}R_B)$, which will 
be very close to 1 given that the entire disk is nearly optically 
thin (the bulge of course is likely to be optically thick and have a 
different value of $k_{\nu}$, but our observations are sensitive to the 
FUV photons that successfully escape from the bulge region so this 
does not affect the results).  If {$L_{\nu}$}/(${\pi}R^{2}2h$)  is
the same for the bulge and disk models,  
the expected values of radiation intensity are equal
to an accuracy within 10 $\%$, 
i.e. $<J_\nu^B> \simeq J_\nu^D$ where $J_{\nu}^{D}$ is
the mean intensity computed for disks with a uniform 
distribution of sources in
equation (14) in Paper I. Since these two models represent 
extremes of the source distribution (uniform versus central source) 
we expect any intermediate source distributions to lead to similar 
values of radiation intensity.  This is important since hierarchical 
structure formation implies that a given DLA could receive radiation 
from several compact regions of active star formation rather than 
a single central bulge. Because   
the SFRs per unit comoving volume for bulges or disks are given by
\begin{equation}
{\dot{\rho_{*}}}{ \ \propto} \ L_{\nu}n_{co}
\cmma
\label{eq:rhodotB}
\end{equation}

\noindent the bulge-to-disk ratio ({\rhodot})$_{B}$/({\rhodot})$_{D}$=
$L_{\nu}^{B}$/$L_{\nu}^{D}$. Comparison between equations
14 in Paper I and equation ({\ref{eq:JnuB_average}}) 
in this paper shows this ratio
equals the product of {\\} $<J_{\nu}^{B}>$/$J_{\nu}^{D}$ and the
ratio of the dimensionless integrals in both equations.
To compare bulge and disk SFRs we assume that 
$J_{\nu}^{D}$=$<J_{\nu}^{B}>$ so that both models generate the
observed heating rate. Because the 
dimensionless integrals 
are identical  to within 10$\%$, we find that ({\rhodot})$_{B}$$\approx$
({\rhodot})$_{D}$.

As a result
the estimates of {\rhodot} deduced for star formation
throughout uniform disks 
do not change significantly when star formation 
is confined to the centers of
such disks. 
Therefore, 
our estimates of {\rhodot} for disks do not appear to be very sensitive
to the radial distribution of the sources of FUV radiation provided
the disks are optically thin to such radiation.
Though the disks
are likely to be optically thin, we cannot rule out the presence
of optically thick dust in the bulge, which attenuates some
fraction of the FUV radiation emitted by the stars. In that case the 
expression for ({\rhodot})$_{B}$ is found by equating {\rhodot}
for the disk and bulge models and the resulting   
({\rhodot})$_{B}$ 
would be a lower limit to the actual
SFR per unit comoving volume.
Since we can increase {\rhodot}($z{\ge}$1.6)
inferred for the consensus model, i.e., inferred
from FUV heating, by a factor of 3.3 before the background
is violated, the radiation attenuated in the bulge model
can be as much as a factor of 2.3 times that observed; i.e.,
as much as 0.7 of the total FUV radiation can be attenuated. 

We can also use the bulge model to compute the {\em total} SFRs, {\ms},
required to explain the observed {\ciis} heating rates.
From equation ({\ref{eq:JnuB}}) we find that {\ms}=
1.9$(r/10 {\rm kpc})^{2}$$G_{0}$  $M_{\odot}$yr$^{-1}$, where we 
used the Madau {\etal} (1996) calibration
to convert $L_{\nu}$ to {\ms}. Assuming $G_{0}$ = 6.8, 
the average of the positive detections for the ``Gal'' 
minimal depletion model, we have {\ms}=13$(r/10 {\rm kpc})^{2}$
$M_{\odot}$yr$^{-1}$. If we let $r$ equal 
10 kpc, a typical impact parameter, we find that {\ms}
is consistent with upper limits
obtained from H$\alpha$ imaging of 7  DLAs by Bunker {\etal} (2001). 
On the other hand this SFR is
about 3 times higher than the more sensitive NICMOS upper limits
reported by Kulkarni {\etal} (2001) for the DLA toward
Q1244$+$34. Unfortunately we do not have
{\ciis} 1335.7 profiles  of this DLA. 
Future searches for emission from DLAs with
detected {\ciis} absorption
may prove to be sensitive
tests of the bulge hypothesis. These tests
will be less sensitive
for the uniform disk models, where
the predicted surface brightnesses are low. 
In any case the limits on {\ms} for the bulge 
hypothesis place severe constraints
on rotating disk models in which $r$$\ge$20 kpc is required (PW97),
since {\ms} for the central bulge will be higher than
observational constraints allow.

\subsection{Search for Evidence of Star Formation and Feedback}

Stars leave imprints on the gas from which they form, and this
may be detectable in DLAs. In the uniform disk scenario 
a correlation should exist between {\ps} and {\\} {\nh} resembling
the Kennicutt  relation found in nearby disk galaxies
(Kennicutt 1998).
This is a manifestation of the  condensation of gas 
into stars, since it is equivalent to the Jeans
instability criterion in rotating disks (Toomre 1964).
Stars also deposit
energy and other byproducts of stellar evolution into
the gas, and so one might expect to find evidence of
feedback if the uniform disk scenario is correct.
This would occur through shocks
generated by supernova explosions. Because of the short main sequence
lifetimes of the progenitor stars, feedback is directly related to 
the SFR. By contrast, in the bulge scenario
the connection between stars and
DLA gas may (a) not exist or (b)  be indirect.

We test these ideas with SFRs derived from the CNM model with ``Gal''
dust and minimal depletion. 
The results of this investigation are
statistically indistinguishable from those of ``Gal'' dust with maximal
depletion, and ``SMC'' dust with maximal or minimal depletion.
The test of the Kennicutt relation is shown in Figure
7a where we plot \\ ({\nh}, {\ps}) pairs for DLAs along with the 1-$\sigma$
contours of the Kennicutt 
relation.\footnote{Although the total H column density
is used for nearby galaxies, 
{\nh} is adequate for DLAs where 
the molecular content of the gas is low (Petitjean {\etal} 2002);
Ledoux {\etal} 2002}
Clearly the DLAs show no evidence for a Kennicutt relation. Specifically,
a Kendall {$\tau$} test using only positive 
detections reveals {$\tau$} = 0.25 and
the probability that the null hypothesis of no correlation is correct is
$p_{Kendall}$=0.28. This conclusion differs
from 
the findings of Lanzetta {\etal} (2002) who
used the Kennicutt relation to deduce the SFR intensity distribution
as a function of SFR per unit area from the frequency distribution
of H I column densities in DLAs.
They 
found 

\begin{figure}[h]
\centering
\scalebox{0.34}[0.39]{\rotatebox{-90}{\includegraphics{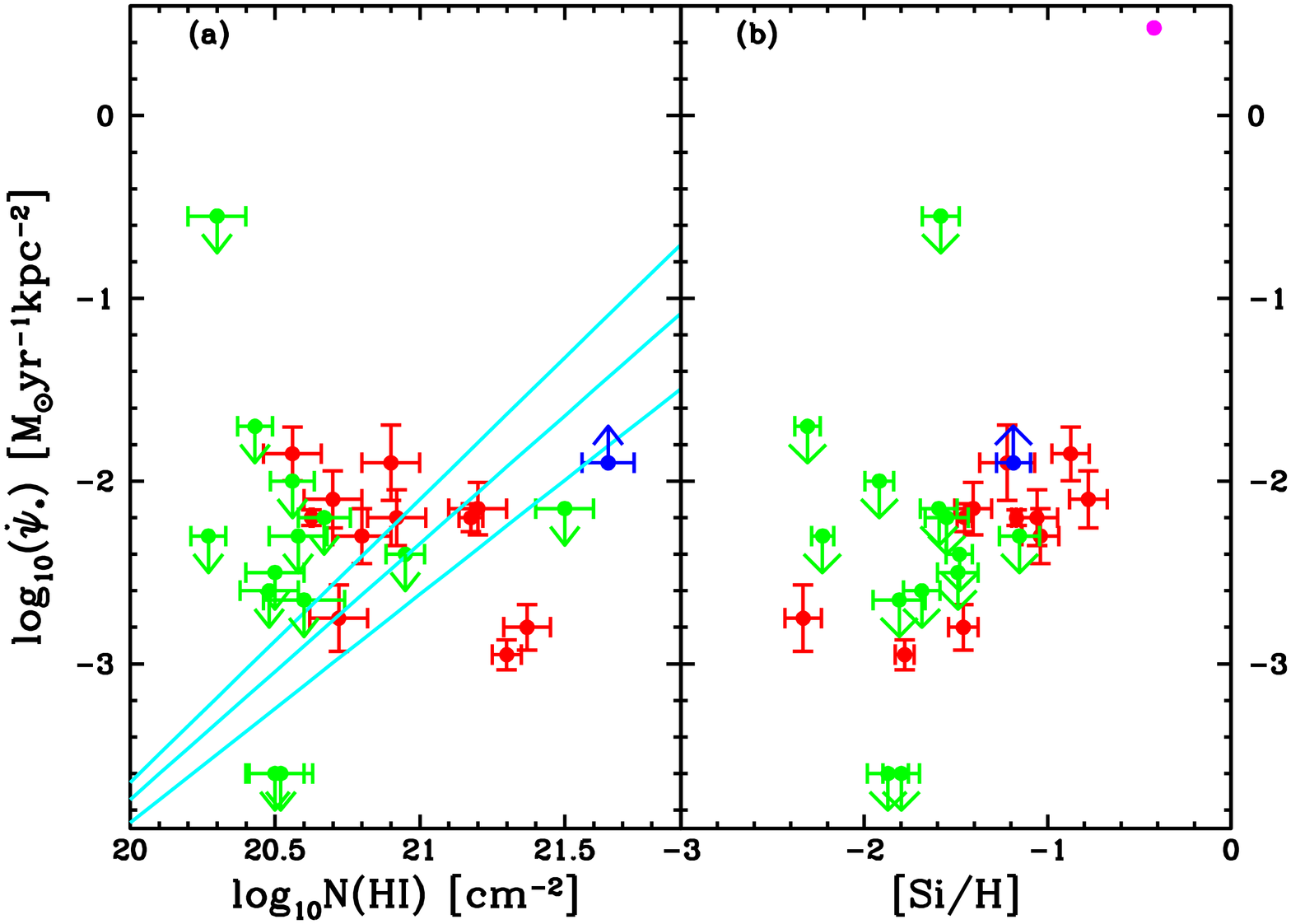}}}
\caption[]{(a) Comparison between {\nh} and {\ps} for DLAs with
Kennicutt (1998) relationship defined by
{\ps} =
(2.5$\pm$0.7{$\times$}10$^{-4}$){$\times$}({\NH}/1.26{$\times$}10$^{20}$ cm$^{-2
}$)$^{1.4{\pm}0.15}$
{\smpykpc}. Data points inferred with ``Gal'' dust and minimal depletion.
Red data points are positive detections, green are
2-$\sigma$ upper limits, and blue is 2-$\sigma$ lower
limit. (b) Comparison between [Si/H] and {\ps}. Same
color coding as 7a. Magenta circle is estimated mean for LBGs.}
\label{rhovszSMC}
\end{figure}

\noindent excellent agreement with the 
intensity distribution deduced directly from
galaxy brightnesses. In that case {\ps} is 
deduced from emission from pixels with linear 
dimension of $\approx$ 0.2$h^{-1}$ 
kpc. By contrast, the H I column densities 
in DLAs are sampled
over transverse distances determined by the linear scale of the continuum
source in QSOs, which is typically less than 1 pc.
As a result, the agreement between column density measurements
on small scales with SFRs per unit area on larger scales indicates
that {\em on average} the Kennicutt relation for nearby galaxies holds in 
high-$z$ DLAs (Lanzetta {\etal} 2002).

The reasons why the points in Figure 7a display so much scatter about
the Kennicutt relation are straightforward. First, if star formation
occurs in the DLA gas, {\ps} inferred from the {\ciis} technique is
averaged over the linear dimensions of the DLA, which exceed 5 kpc in
any model (e.g. Haehnelt {\etal} 1998).  Because {\NH} likely varies
on scales smaller than 5 kpc, correlations between ({\NH},{\ps}) pairs
are not expected.  Second, in the bulge model, star formation does not
occur in the gas giving rise to damped {\lya} absorption. As a result
no correlations between ({\NH},{\ps}) pairs are predicted.  However,
it may be possible to distinguish between the two models with a
sufficiently large data set. If star formation occurs in the DLA gas,
{\NH} averaged over the DLA sample should correspond to {\NH} averaged
across a typical DLA, and therefore the sample averages of {\NH} and
{\ps} should obey the Kennicutt relation.  On the other hand, no such
correlation is predicted for the bulge model. Interestingly, the
averages over the positive detections in Figure 7a result in
{\psav}=6.4{$\times$10$^{-3}$} {\smpykpc}
and {$<{N({\rm H I})}>$}=1.0$\times$10$^{21}$ cm$^{-2}$,
which are within 1$\sigma$ of 
the

\begin{figure}[h]
\centering
\scalebox{0.34}[0.39]{\rotatebox{-90}{\includegraphics{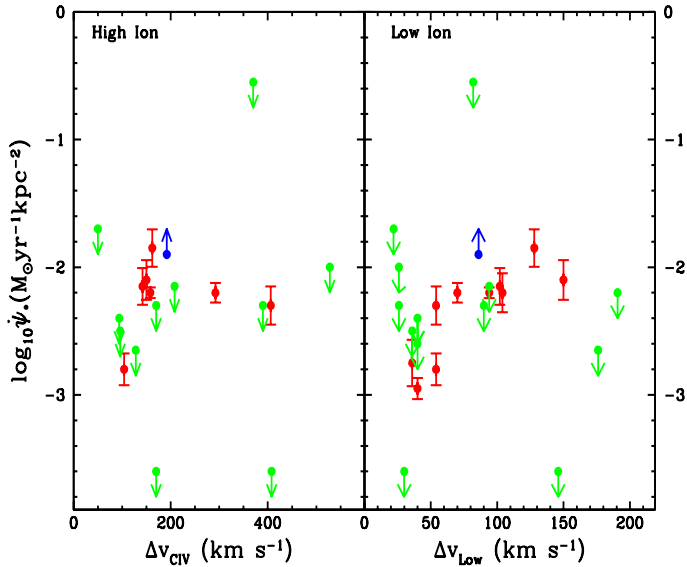}}}
\caption[]{(a) Comparison between C IV velocity width
${\Delta v}_{C IV}$  and {\ps}. (b) Comparison between low-ion
velocity width ${\Delta v}_{Low}$ and {\ps}.
Color coding same as in previous figures. }
\label{deltavvspsi}
\label{rhovszSMC}
\end{figure}

\noindent average Kennicutt relation. In the bulge model this is merely a
coincidence.

To search for evidence of feedback, we first looked for
correlations between
SFR per unit area and metallicity. Nearby spirals exhibit negative radial 
gradients in metallicity (Garnett {\etal} 1997) and in SFR per unit
area (Dopita \& Ryder 1994), implying a correlation between metallicity
and SFR per unit area. Such metallicity gradients may have
also been detected in DLAs (Wolfe \& Prochaska 1998). We used ([Si/H], {\ps})
pairs to search for such correlations. The results shown in Figure 7b
yield tentative evidence for a correlation since 
{$\tau$}=0.46 and $p_{Kendall}$=0.05,
where again we used only positive detections.
We  then focused on kinematic evidence for feedback. In the ISM,
enhancements in velocity width are found in regions of higher than
average SFRs such as Orion (Cowie {\etal} 1979) and Carina (Walborn
{\etal} 1998; Savage {\etal} 2001). Evidently, gas in these regions
is stirred up by increased supernova activity. We tested for kinematic
feedback by looking for correlations between absorption-line
velocity width
and {\ps}. Specifically, we checked for kinematic feedback in the
neutral gas by searching for correlations between 
(${\Delta v}_{Low}$, {\ps}) pairs, where {${\Delta v}_{Low}$} is 
the velocity width of low ions in DLAs (Prochaska \& Wolfe 1997; hereafter
PW97).
We also checked for kinematic feedback in the ionized gas by
searching for correlations between 
(${\Delta v}_{C IV}$,{\ps}) pairs, where {${\Delta v}_{C IV}$} is
the velocity width of the  C IV 1550 transition.
The results using only positive detections are shown in Figure 8a
and  reveal no evidence for correlations
in the ionized gas, since $\tau$=0.97 and $p_{Kendall}$=0.76. But 
Figure 8b does reveal possible
evidence for kinematic correlations in the neutral gas, since
$\tau$=0.77 and $p_{Kendall}$=0.002.

The reasons for null correlations in the case of feedback are the
same as discussed above. Namely, no correlations are expected
for the bulge hypothesis because star formation occurs in regions 
disconnected from the absorbing gas used to infer metallicity
and kinematics. 
In the disk model, {\ps} is averaged over
linear scales large compared to the {$\approx$} l pc transverse dimensions
sampled in absorption. Therefore,
in this model the null correlation between 
${\Delta v}_{C IV}$ and {\ps} indicates significant
random variations in high-ion velocities on
linear scales small compared to the length scale of the 
star forming regions in DLAs.
Because it is reasonable to expect similar variations
in the case of low-ion velocities, how can we understand
a correlation between the 
(${\Delta v}_{Low}$, {\ps}) pairs,
if confirmed? The answer may be that ${\Delta v}_{Low}$ has
global rather than local significance. Specifically,
${\Delta v}_{Low}$ may reflect the depth of the
gravitational potential well of the DLA, as predicted in the
case of rotating disks (PW97) or protogalactic
clumps (Haehnelt {\etal} 1998); i.e., the SFR per unit area may be
correlated with total mass. Note, this explanation would
apply both to the uniform disk and bulge models. A similar explanation
might also apply if the tentative correlation between ([Si/H],{\ps})
pairs is confirmed; namely, that the metallicity of the gas does not
vary randomly on scales small compared to scale of the 
galaxy hosting the DLA. Rather,
metallicity might be a function of total mass, as in the case 
of current galaxies.

Of course all these
results need to be tested with more data.
In particular, the statistical significance of the correlation 
between ${\Delta v}_{Low}$ and {\ps}
would be reduced if  the two upper limits with 
$>$ 150 {\kms} were added to the sample of positive detections.
On the other hand, 
the two DLAs with
log$_{10}${\ps}$\approx$$-$3.6 {\smpykpc} are the ``outliers''
discussed in Paper I. These are more likely to be WNM-dominated absorbers
with significantly higher values of {\ps}, in agreement 
with the predicted correlations.

\section{TESTS OF THE MODELS}
Having presented evidence for star formation in DLAs, and
having the described the implications of the derived star formation
histories, we now
discuss three tests of the models upon which these
results are based. The first is a test for the grain photoelectric
heating mechanism, the second tests the hypothesis that
{\lclos} is a cooling rate, and the third describes a 
search for evidence of a two-phase medium.

\subsection{Evidence For Grain Photoelectric Heating}

A critical
premise of the {\ciis} technique
is that neutral gas in DLAs is heated
by photoelectrons ejected from interstellar grains by FUV radiation
emitted by massive stars.
In that case 
the heating rate per H atom, {\gamdnr} {$\propto$}
{\kapnr}{$\epsilon$}{\ps}. 
The efficiency of grain photoelectric heating,
{$\epsilon$}, is essentially constant as it is insensitive
to variations in electron density, temperature, and FUV
radiation field in the CNM.
Moreover,
the scatter in {\ps} is limited 
such that the average 
log$_{10}${\psav}=$-$2.19$^{+0.19}_{-0.26}$ {\smpykpc}.
Therefore,
a prediction of the grain photoelectric heating scenario
is that 
{$\Gamma_{d}$} should
be roughly correlated
with the dust-to-gas ratio, {\kapnr}.

\begin{figure}[ht]
\includegraphics[height=3.8in, width=3.6in]{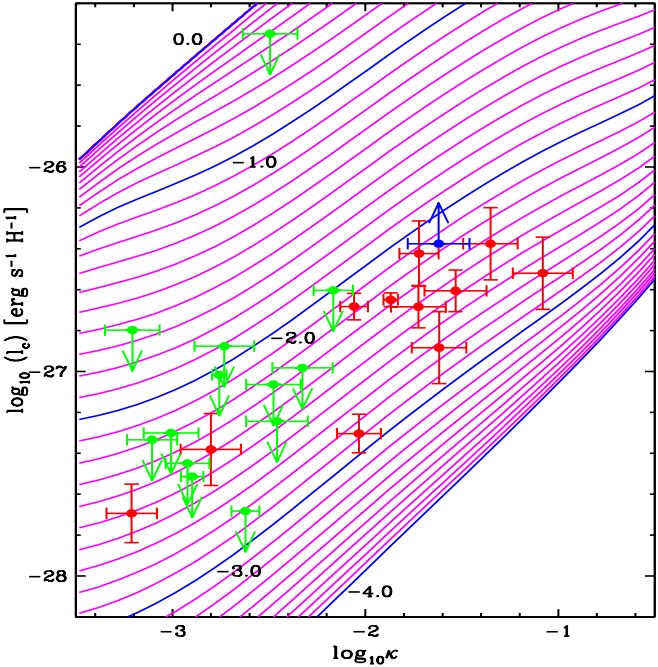}
\caption{Plot of dust-to-gas ratio, {\kapnr} versus {\lclos}
where the ``Gal'' dust and minimal depletion is assumed to compute
{\kapnr}. Color code of data points same as in previous figures.
Continuous curves  are {\kapnr} versus {\lcrnr} relations
for constant log$_{10}${\ps} = $-$4.0, $-$3.0, $-$2.0, $-$1.0,
and 0.0 {\smpykpc} predicted by
model described in text.}
\label{metals density}
\end{figure}

We test the grain photoelectric heating hypothesis by letting {\lclos}={\gamdnr}
(which is an excellent approximation in the CNM)
and then
compare {\lclos} with {\kapnr}. The two quantites are plotted
against each other in Figure 9. The red data points are
positive detections, green are upper limits, and blue
is a lower limit.
The solid curves are lines of constant {\ps} predicted
by a version of the CNM model with a fixed redshift
and a given prescription for computing {\kapnr} from metallicity.
That is, for a given metallicity and {\ps} we calculate two-phase
equilibria of gas subjected to grain photoelectric heating and
assume the DLA density to be given by the computed $n_{CNM}$.
We include the CMB contribution to radiative excitations
of the [C II] fine-structure states
by assuming $z$ = 2.8, the 
median redshift of the sample, and ignore optical 
pumping (which should lead to no loss in generality since
optical pumping is negligible in the CNM). To
calculate {\kapnr} we adopt the CNM ``Gal'' model with minimal depletion.
As a result we let the intrinsic carbon abundance,
[C/H]$_{int}$=[Si/H]$_{int}$$+$[Fe/Si]$_{int}$ and
[Fe/Si]$_{int}$ = $-$0.2.
We also assume [Fe/Si]$_{gas}$=$-$0.4,
which is the  average value for our sample. 
We then compute {\lcrnr} with techniques described in 
Paper I.
Visual inspection of 
Figure 9 indicates several phenomena. First, the measured 
{\lclos} and {\kapnr} are correlated. Specifically, performing
the Kendall tau test using the positive detections alone we find
$\tau$=0.64 and  $p_{Kendall}$=0.0064,
where $p_{Kendall}$ is the probability of the null hypothesis
of no correlation. Second, the slope of the
data in the ({\kapnr}, {\lclos}) plane is 
approximately parallel to the model predictions,
and all the positive detections are bounded
by log$_{10}${\ps} = $-$2 and $-$3 {\smpykpc}.
Third, 
the case for correlation receives
additional support from 
the location of the upper limits at low dust-to-gas ratios,
i.e., log$_{10}${\kapnr} $<$ $-$2.2, and the lower limit at
the relatively  high dust-to-gas ratio,
log$_{10}${\kapnr} $>$ $-$1.6.  

Can the correlation between {\lclos} and {\kapnr} be explained by
heating mechanisms other than grain photoelectric
heating? The obvious alternatives are cosmic-ray
and X-ray heating.
Suppose the actual SFRs are orders of magnitude lower than
we infer for grain photoelectric heating, but the cosmic-ray ionization
rate, {$\zeta_{CR}$}, is large; i.e., the ratio {$\zeta_{CR}$}/{\ps}
is orders of magnitude larger than given in equation (9) in
Paper I. In that case it
is possible for the cosmic-ray heating rate, $\Gamma_{CR}$,
to dominate the heating rate in
the CNM. During ionization by cosmic rays, primary and secondary electrons
are mainly liberated from H and He
(W95).  As a result,
$\Gamma_{CR}$ will be independent of metal abundance, hence independent
of {\kapnr}. 
Since we have linked {$\zeta_{CR}$}, hence {$\Gamma_{CR}$},
to the SFR, and {\lclos}$\approx$$\Gamma_{CR}$
when cosmic rays dominate,
{\lclos} will be independent of {\kapnr} for a fixed {\ps}, in contrast to 
grain photo-electric heating.

On the other hand, the
X-ray heating rate could depend on metallicity, hence on {\kapnr},
since abundant heavy elements dominate the X-ray photoionization
cross-section per H atom at photon energies above the Oxygen edge
at 0.53 keV (Morrison \& McCammon 1983). 
We used the  X-ray heating model
of W95, which
consists of two local thermal sources
with $T$ $\sim$ 10$^{6}$ K and an extragalactic power-law component.
Heating by the thermal sources is dominated by photoionization of
H and He since their X-ray spectra cutoff below 0.53 keV.
Although the power-law component extends to energies above the
Oxygen edge, X-ray heating in this case will also be dominated by photoionization
of H and He. The reason is that for typical velocity-component column
densities, {\NH} $\approx$ 1$\times$10$^{20}$ cm$^{-2}$, most of
the X-rays penetrating the H I gas have energies below 0.53 keV. 
Photoionization of H and He by 
these X-rays dominates photoionization of heavy elements by higher energy 
X-rays due to the 
low metallicities of most DLAs and the shape of the power-law spectrum.
Thus, photoionization of heavy elements
will not be the dominant source of primary electrons.
Consequently, while the condition {\lclos} $\approx$ {$\Gamma_{XR}$} might 
hold
in the limit of low SFRs and high X-ray luminosities,
{$\Gamma_{XR}$} will be independent of {\kapnr} for a fixed {\ps}.

To conclude, the correlation between {\lclos}
and {\kapnr} is naturally explained by grain photoelectric heating.
The correlation at a fixed SFR follows from the physics
of grain photoelectric heating, while the scatter of \ {\lclos} at
fixed {\kapnr} reflects the frequency distribution
of  {\ps}. Note, this correlation does not distinguish between
the uniform disk and bulge models. Because the heating
rate is the product of a ``global'' quantity, {\ps}, and a ``local''
quantity, {\kapnr}, clouds with larger {\kapnr}
will have higher heating rates independent of whether the
incident FUV radiation arises locally or from the distant bulge.
By contrast, neither the cosmic-ray nor X-ray heating rates
are correlated with {\kapnr} at fixed SFR. To explain
the observed correlation, one must postulate a correlation between
{\ps} and {\kapnr} (or possibly metallicity).
It is difficult to understand the origins of a correlation
between the ``global'' 
{\ps} and the ``local'' dust-to-gas ratio in this case. 

\subsection{Is {\ciis} Excitation Due to the CMB?}

Our models predict that in the CNM of most DLAs the cooling rate,
$n{\Lambda}$, equals the spontaneous emission rate of 158 {\micron}
radiation, {\lclos}. This condition, $n{\Lambda}$$\approx${\lclos}, holds
when the $^{2}P_{3/2}$ and $^{2}P_{1/2}$ fine-structure states in the
ground term of C$^{+}$ are populated by collisional rather than

\begin{figure}[h]
\centering
\scalebox{0.36}[0.46]{\rotatebox{-90}{\includegraphics{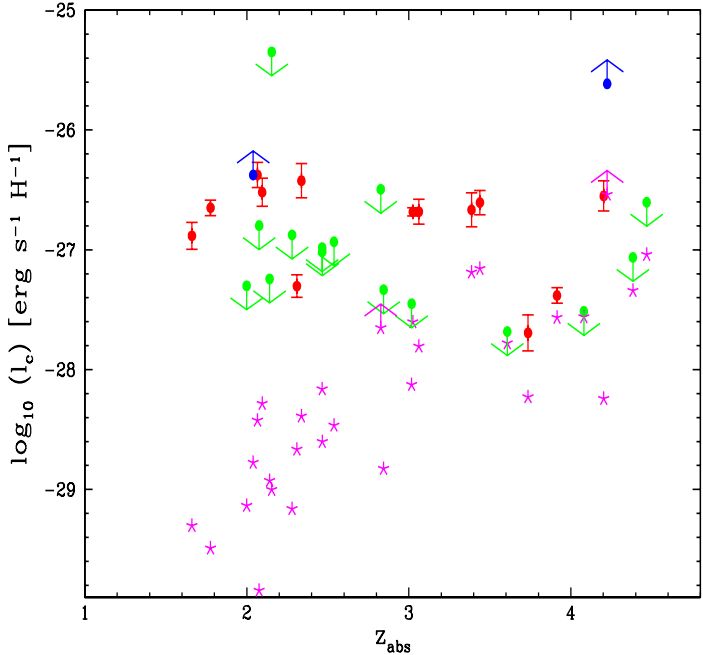}}}
\caption[]{Plot of {\lclos} versus redshift. Red, green, and blue
filled circles have usual meanings. Magenta stars depict
({\lclos})$_{CMB}$ corresponding to measurements of {\lclos}
at each redshift. The plot shows data for 30 DLAs. Computation
of ({\lclos})$_{CMB}$ explained in text.}
\label{rhovszSMC}
\end{figure}

\noindent radiative excitations. In Paper I we described how CMB radiation 
populates these states directly, and how the FUV 
radiation field,
$G_{0}$, populates them indirectly through optical pumping via
higher energy states. While optical pumping is important in the
WNM, it can be neglected in the CNM. Although an increase in
$G_{0}$ drives up the pumping rate, it also increases
the collisional excitation rate. This is because an increase in
$G_{0}$, and thus {\ps}, increases the grain photoelectric heating rate, which
raises the CNM density, $n_{CNM}$, as shown in Figure 5
in Paper I. As
a result, the ratio of collisional to optical pumping
rates always exceeds unity in the CNM.
\footnote{Although the density
of the WNM, $n_{WNM}$, also
increases with $G_{0}$, the densities are not high
enough for collisions to dominate optical pumping
in most cases.}
This need not be true for the ratio of collisional to  CMB
excitation rates   because the CMB does not
heat the CNM gas (through Compton heating) in the redshift range
of our sample DLAs. Because the value of 
$n_{CNM}$ does not rise with increasing CMB intensity,
the 
CMB may dominate collisions as a source of excitation in 
DLAs with low values of $n_{CNM}$ and high redshifts where the
CMB intensity is high.

Because the balance between collisional and CMB excitations
depends on various assumptions included in our models, it
is important to assess their relative importance with model-independent
tests.
Figure 10 illustrates the
results of a test 
relying on one free parameter, the carbon abundance, (C/H).
Here we plot {\lclos}, inferred from measurements of
$N$({\ciis}), versus redshift for a sample
of 30 DLAs. For each DLA we also plot, as magenta stars,
({\lclos})$_{CMB}$ versus redshift
where ({\lclos})$_{CMB}$ is the value assumed by {\lcrnr} when the
CMB is the only source of excitation and de-excitation (in Paper I
we showed that  
$(l_{cr})_{CMB}=2({\rm C/H})A_{ul}h{\nu}_{ul}{\rm exp}{\biggl [}-h{\nu}_{ul}/{\bigl [}k(1+z)T_{CMB}{\bigr
 ]}{\biggr ]}$, 
where  the present CMB temperature, $T_{CMB}$ 
=2.728 K.
Twenty five of the DLAs are drawn from the minimal depletion
sample. Here we calculate ({\lclos})$_{CMB}$ by assuming
[C/H]=[Si/H]$-$0.2 to compute (C/H).
The plot also shows data for 5 additional DLAs. They include the DLAs
toward Q0201$+$11 and Q2344$+$12 for which we assumed maximal depletion
to obtain (C/H) because [Fe/Si] $>$ $-$0.2 in both DLAs. We also
included the DLAs toward Q0951$-$04, Q1425$+$60, and Q1443$+$27 for which
we assumed [C/H]=[Si/H]$-$0.2. These objects were excluded from the original
minimal depletion sample because observational limits on [Si/H] or 
[Fe/H] prevented us from computing the dust-to-gas ratio, {\kapnr}, which
is not needed to compute ({\lclos})$_{CMB}$.
\noindent The figure demonstrates that 
the CMB cannot explain the observed level of {\ciis} excitation
for the following reasons: First, despite the dispersion in carbon
abundance, 
({\lclos})$_{CMB}$ should increase rapidly
with increasing redshift whereas
the observed values of {\lclos} show
no dependence on redshift in the interval $z$ = [1.6, 4.5]. Second,
with two  exceptions, the CMB excitation rate is 
too low to explain the observations. This is true even at $z$ $>$ 3 where
the predicted values of ({\lclos})$_{CMB}$ merge
with the observed spontaneous emission rates.

The exceptions are
the DLAs at $z$=3.608 and $z$=4.080 toward Q1108$-$07 and  Q2237$-$06,
the ``outliers'' discussed in Paper I. In these DLAs,
({\lclos})$_{CMB}$ comprises a significantly higher fraction 
($>$ 0.8) of {\lclos} than for the other DLAs. Because the {\lclos}'s are upper
limits, the data place lower than usual upper limits on the
[C II] 158 {\micron} cooling rates. We interpreted
this to mean that these sightlines pass only through WNM gas subjected
to SFRs within the range determined from the CNM
hypothesis. However, we cannot rule out the possibility that
the gas is subjected to negligible SFRs, which lead to
gas densities so low that collisional excitations are
unimportant. In that event, CMB excitation
alone would be responsible for the observed {\lclos}. To decide between
these hypotheses, we need to measure {\lclos} and ({\lclos})$_{CMB}$
more accurately. Though {\lclos} can be determined more
precisely through better measurements of the {\ciis} 1335.7 velocity
profiles, higher signal-to-noise will not
improve the accuracy of  ({\lclos})$_{CMB}$ where the
only source of error is in determining (C/H). The carbon
abundance is difficult to obtain directly because the
principal C II transitions are always saturated
(e.g. Dessauges-Zavadsky {\etal} 2001). Instead we compute
({\lclos})$_{CMB}$ with the minimal depletion assumption,
[C/H]=[Si/H]$+$[Fe/Si]$_{int}$ and [Fe/Si]$_{int}$=$-$0.2,
for 28 of the 30 data points shown in 
Figure 11. For the remaining 2 DLAs,
[Fe/Si]$_{gas}$
$>$ $-$0.2, which violates the nucleosynthetic
ceiling limit in the minimal depletion model, and we thus assume
[Fe/Si]$_{int}$=0; i.e., maximal depletion.
We favor minimal depletion for most DLAs because it leads to
self-consistent determinations of ({\lclos})$_{CMB}$, whereas
maximal depletion leads to the inconsistent condition
({\lclos})$_{CMB}$ $>$ {\lclos} for the Q1108$-$07 and Q2237$-$06
DLAs. In any case the limited accuracy for determining
(C/H) results in errors in ({\lclos})$_{CMB}$ of order
0.2 dex, which is inadequate for distinguishing 
collisional excitation  from CMB excitation.

In our treatment of CMB excitations of the C II fine-structure states 
we assume the CMB temperature at redshift, $z$,
$T_{CMB}(z)=(1+z)T_{CMB}$, whereas most published

\begin{table*}[ht] \footnotesize 
\begin{center}
\caption{{\sc PHYSICAL PARAMETERS FOR DLA GAS}}
\begin{tabular}{lcccc}
\tableline
\tableline
\cline{2-5}
&\multicolumn{2}{c}{Q1232$+$08}&\multicolumn{2}{c}{Q0347$-$38}  \\
\cline{2-3} \cline{4-5}
Parameter &Estimate$^{a}$&CNM Model$^{b}$&Estimate$^{c}$&CNM Model$^{b}$   \\
\tableline
$n$(cm$^{-3}$) & 20$-$335&7$-$14&4$-$14&3$-$6  \\
$T$ (K)& 85$-$285&130$-$188&$<$950$^{d}$&130$-$200  \\
$n_{e}$(cm$^{-3}$) & 0.02&0.002$-$0.03&...&0.002$-$0.014  \\
$G_{0}$ &$\sim$1.7$^{e}$&6.0$-$21.5&$\sim$ 1.7$^{e}$&2.7$-$11.0 \\
\tableline
\end{tabular}
\end{center}
\tablenotetext{a}{From Srianand {\etal} (2000).}
\tablenotetext{b}{Parameter range predicted by CNM models with ``Gal'' and ``SMC''
dust and minimal and maximal depletion.}
\tablenotetext{c}{From Levshakov {\etal} (2002).}
\tablenotetext{d}{Upper limit from velocity dispersion $\sigma$ = 1.4 {\kms}
observed for H$_{2}$ lines differs from estimate of Levshakov {\etal} 2002.}
\tablenotetext{e}{Inferred from reported photoabsorption rate in Lyman and
Werner bands, ${\beta}_{0}$}
\end{table*}

\noindent {\em test} this assumption by  attempting
to measure
$T_{CMB}(z)$ directly. In principle this can be done by first making independent
estimates of the collisional contribution to the level populations
of the $^{2}P_{3/2}$ and $^{2}P_{1/s}$  fine-structure states,
and then computing the black-body temperature required to explain
the observed population ratio (e.g. Molaro {\etal} 2001).
Accurate UVES echelle spectra
were acquired by the VLT for two DLAs in our sample; Q0347$-$38 (Levshakov
{\etal} 2001; Molaro {\etal} 2001)
and Q1232$+$08 (Srianand {\etal} 2000). In Table 2 we compare physical parameters
obtained by these authors
with predictions of
our CNM models. 
The estimates are independent of our predictions because
physical models of the DLA gas were not constructed. Rather
techniques such as measuring velocity line widths to
estimate the kinetic temperature or inferring $G_{0}$ from 
the fractional abundance of H$_{2}$ were used. In some cases
relative abundances of various rotational levels of H$_{2}$ were
used to infer the temperature, and the
ratio Mg$^{+}$/Mg$^{0}$ was combined with
$G_{0}$ and photoionization equilibrium to
obtain $n_{e}$. The results in Table 2
show, with the possible exception of $G_{0}$, reasonable agreement between our predictions and these estimates.
They are also inconsistent with the WNM hypothesis for both DLAs.

\subsection{Ratios of {\ciis} to Resonance-line Velocity Profiles:
Probing the Two-Phase Medium}

Here we discuss an observational test of the two-phase medium,
a key element
in our models of DLAs. We describe various aspects of the test
and leave more quantitative analyses  for future discussions.  
Note, Lane {\etal} (2000) provide independent evidence for a 
two-phase medium from their analysis of 21 cm absorption in
a DLA with $z$ = 0.0912.

Suppose, by analogy with the ISM, the probabilities that the line of sight
intercepts WNM and CNM clouds
are comparable (Kulkarni  \& Heiles 1987). If
velocity components (i.e., clouds) in each phase have comparable H I
column densities and the same element abundances, then a random
sightline through a typical DLA should encounter similar column
densities of metals in each phase. This can lead to measurable
differences between the velocity profiles of {\ciis} 1335.7
and resonance transitions such as Si II 1808.0, as we now show.

Consider a two-phase configuration in which the
fractions $y_{CNM}$ and $y_{WNM}$ of the total {\NH} are in the
CNM and  WNM, where $y_{CMN}$$+$$y_{WNM}$=1.
Assume the systemic cloud velocities to be $u_{CNM}$ and $u_{WNM}$, and the
internal Gaussian velocity dispersions to be ${\sigma}_{CNM}$
and ${\sigma}_{WNM}$. In that case
the optical depths at velocity
$v$ in {\\} {\ciis} 1335.7 and Si II 1808.0  are given by:

\begin{equation}
{\tau}_{v} = {{{\pi}e^{2}} \over {mc}}{1 \over {\sqrt {2{\pi}}}}N({\rm H I})\left\{
\begin{array}{ll}
f_{{\rm C II}^{*}}{\lambda}_{{\rm C II}^{*}}{(l_{c})_{CNM} \over {A_{ul}h{\nu}_{ul}}}{\biggl[}y_{CNM}{{\Phi}(v-u_{CNM}) \over {\sigma}_{CNM}} \\
+y_{WNM}{{(l_{c})_{WNM}} \over (l_{c})_{CNM}}{{\Phi}(v-u_{WNM}) \over {\sigma}_{WNM}}{\biggr]} \ ; \ {\rm C II}^{*} \\
f_{{\rm SiII}}{\lambda}_{{\rm SiII}}({\rm Si/H}){\biggl[}y_{CNM}{{\Phi}(v-u_{CNM})\over {\sigma}_{CNM}} \\
+y_{WNM}{{\Phi}(v-u_{WNM}) \over {\sigma}_{WNM}}{\biggr]}\ ; \ {\rm Si II} 
\cmma
\end{array}
\right.
\label{eq:tauvciistar}
\end{equation}

\noindent where we used equation (3) in Paper I to compute {\\}
$\tau_{v}$({\ciis}),
the  $f$'s 
and $\lambda$'s are oscillator strengths
and 
transition wavelengths,
({\lclos})$_{CNM}$ and ({\lclos})$_{WNM}$ are {\lclos} in the CNM and WNM, 
$\Phi$($v$) is the velocity profile normalized such
that $\int{\Phi}(v)d(v/{\sqrt {2{\pi}}}{\sigma})$=1,
and (Si/H) is the Si abundance. 

To compute ({\lclos})$_{CNM}$ and
({\lclos})$_{WNM}$, we evaluate {\lclos} at the phase densities
$n_{CNM}$ and $n_{WNM}$. 
Equation (11) in Paper I shows that 
$l_{c}= n{\Lambda}_{\rm C II}+(l_{c})_{pump}+(l_{c})_{CMB}$    
where $n{\Lambda_{\rm C II}}$ is the net
loss of thermal energy due to 158 {\micron} emission and 
({\lclos})$_{pump}$ and ({\lclos})$_{CMB}$
are the spontaneous energy emission rates in the limits of
pure optical pumping and CMB excitation, which are defined
in equation (12) in Paper I.
At the high
values of $n_{CNM}$, the fine-structure states are mainly populated
by collisions and as a result {\lclos} equals the cooling rate
$n{\Lambda}_{{\rm C II}}$, which equals the grain photoelectric
heating rate in the CNM; i.e.,
\begin{equation}
\begin{array}{ll}
(l_{c})_{CNM}={\Gamma_{d}} \
\perd
\end{array}
\label{eq:lcCNM}
\end{equation}

\noindent But at the 
low values of $n_{WNM}$ any one of the of the 3 terms 
in the above equation for {\lclos}  can dominate. For example
({\lclos})$_{WNM}$$\approx$({\lclos})$_{CMB}$ 
in the $z$=3.736 DLA toward Q1346$-$03 for the CNM solution in which
$G_{0}$=1.95. By contrast
({\lclos})$_{WNM}$$\approx$({\lclos})$_{pump}$ 
in the $z$=2.039 DLA toward Q0458$-$02 for the CNM solution in which
$G_{0}$$>$13.5  (see Figure 5 in Paper I). Without
optical pumping,
({\lclos})$_{WNM}$ {\\}
$\approx$$n{\Lambda}_{\rm C II}$ if the
metallicity is low and observed {\lclos} is high. In this case
{\lclos} does not equal {\gamdnr} because [C II] emission
is {\em not}  the dominant coolant.
If ({\lclos})$_{WNM}$ is governed
by optical pumping, 
the ratio ({\lclos})$_{WNM}$/({\lclos})$_{CNM}$
is rather well determined because the quantities 
{\gamdnr} and $\Gamma_{ul}$ are both proportional
to the radiation intensity $G_{0}$ and to metallicity. 
Combining the last equation with
equations (1) and (7) in Paper I
and assuming [C/H]$_{gas}$=[Si/H]$+$[Fe/Si]$_{int}$ 
we find
\begin{equation}
{(l_{c})_{WNM} \over ({l_{c}})_{CNM}} = 7.5{\times}10^{9}{{({\rm C/H})_{\odot}({\Gamma}_{lu})_{ISM}} \over {{\bigl [}1-10^{[{\rm Fe/Si}]_{gas}-[{\rm Fe/Si}]_{int}}{\bigr ]{\epsilon}_{CNM}}}}
\cmma
\label{eq:lcCNMovWNM}
\end{equation}

\noindent where
$(\Gamma_{lu})_{ISM}$, the optical pumping rate corresponding to the Draine (1978)
radiation field, equals 1.57{$\times$}10$^{-10}$ s$^{-1}$ (Silva \& Viegas 2002),
and $\epsilon_{CNM}$ is the heating
efficiency in the CNM. 
In deriving the last equation we assume the pumping rate corresponding to 
intensity $G_{0}$ is given by $\Gamma_{lu}$($G_{0}$)=$(\Gamma_{lu})_{ISM}$($G_{0}$/1.7). 
Averaging over
the positive detections in  the ``Gal'' minimal depletion
model we find the average ratio $<$({\lclos})$_{WNM}$/({\lclos})$_{CNM}$$>$
=0.097$\pm$0.020 with optical pumping and 
$<$({\lclos})$_{WNM}$/({\lclos})$_{CNM}$$>$
=0.067$\pm$0.011 without pumping.
As a result,
equation ({\ref{eq:tauvciistar}}) implies the 
WNM is undetectable
in {\ciis} 1335.7 for any value of $y_{WNM}$, but is 
detectable in resonance transitions such as 
Si II 1808.0 for $y_{WNM}$ $>$ 0.1. Consequently, a significant decrease in 
${\tau}_{v}$({\ciis})/${\tau}_{v}$(Si II) from the mean  would signify the presence of
the WNM.

Such variations may have been detected in the DLA toward 
Q0347$-$38. Though
Figure 11a shows clear evidence for distinct velocity components 
at $v$=$-$8 {\kms} and $v$=$+$12 {\kms} in Si II 1808
and Fe II 1608, the $v$=$-$8 {\kms} component is  not seen in {\ciis} 1335.7. Rather,
an asymmetric blue wing extends from $v$=$+$12 {\kms} to
$v$=$-$20 {kms}. Because some of this absorption is due
to weak {\ciis} 1335.66 ($f_{1335.66}$/$f_{1335.71}$=0.11), the observed optical depth of the wing
places an upper limit on ${\tau}_{v}$({\ciis} 1335.71). Therefore,
we conjecture that the $v$=$-$8 {\kms}
component consists of WNM gas, while the $v$=$+$12 {\kms}
component consists of CNM gas. The presence of H$_{2}$ absorption
only at $v$=$+$12 {\kms} and the lower S$^{+}$/N$^{0}$
ratio in this component (Levshakov {\etal} 2002) supports this
interpretation. Notice that the relative 
optical depths of these two components in  Fe II 1608 is different than in  Si II 1808.
The most likely explanation is enhanced Fe depletion
in the
$v$=$+$12 {\kms} component. This does not affect
our interpretation of the {\ciis} profile because
enhanced
depletion at $v$=$+$12 {\kms}
cannot explain the weakness of  {\ciis} 1335.7 at $v$=$-$8 {\kms}.
In any case the significant decrease in the ratio, ${{\cal R}({v})}$$\equiv$
$\tau_{v}$({\ciis})/$\tau_{v}$(SiII),
with decreasing velocity shown in panel 4 of Figure 11a is naturally explained
by the presence of WNM gas at negative velocities.

Figure 11b compares {\ciis} and resonance-line profiles for the
DLA toward Q0100$+$13 (PHL 957). By contrast with Figure 11a, the
{\ciis} profile in this case closely resembles the
Si II 1808 and Ni II 1741 profiles. The only

\begin{figure}[h]
\centering
\scalebox{0.34}[0.45]{\rotatebox{-90}{\includegraphics{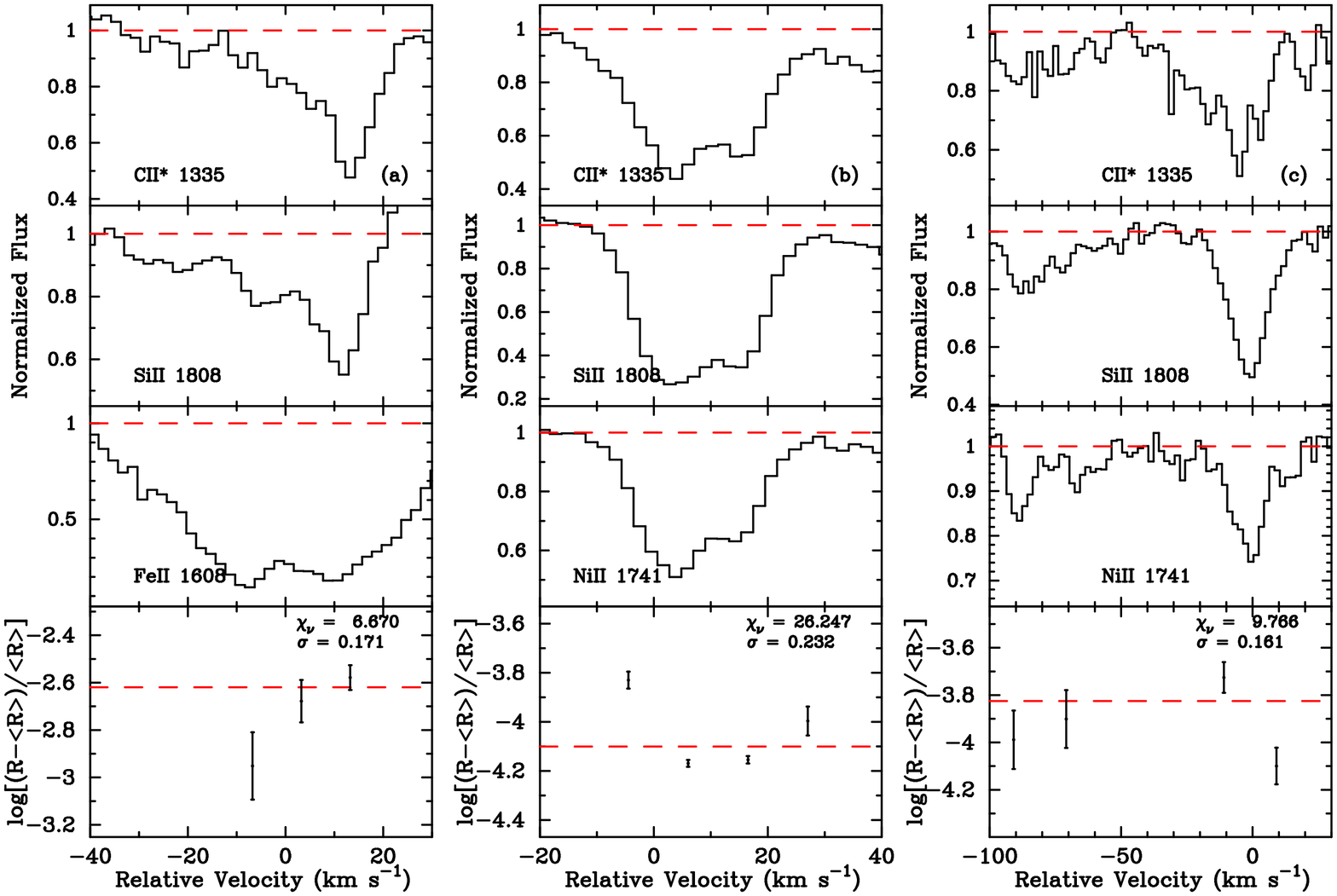}}}
\caption[]{Comparison between {\ciis} and resonance-line
velocity profiles for 3 DLAs.
Top  row of panels show profiles
for {\ciis} 1335.7.
Second row shows profiles for Si II 1808, and
third row for Fe II 1608 (a) and Ni II 1741 (b and c).
Bottom row shows log$_{10}$[(${\cal R}$$-$$<$${\cal R}$$>$)/$<$${\cal R}$$>$]
computed for successive 20 {\kms} bins where ${\cal R}$ is
ratio of {\ciis } 1335.7 to resonance-line optical depth. Columns (a), (b), and
(c) show data for DLAs toward Q0347$-$08, Q0100$+$13, and Q2231$-$00.}
\label{c2starratio}
\end{figure}

\noindent significant difference
is at $v$ $<$$-$8 {\kms} where {\ciis} 1335.71 exhibits a blue
asymmetric wing that is missing from Si II 1808 and Ni II 1741.
We believe this is a blend with 
weak $v$ = 0 absorption in {\ciis} 1335.66. Because we ignore the
increased {\ciis} absorption at $v$ $>$$+$30 {\kms}, which is
likely to be a blend with {\lya} forest absorption lines,
we find no compelling evidence for a WNM 
in this DLA. However, this does not rule out the presence
of a WNM. Suppose the velocity components
at $v$ $\approx$ $+$3 {\kms} and $+$15 {\kms}
each contain WNM and CNM gas with the same velocity dispersion. From 
equation ({\ref{eq:tauvciistar}}) we see the ratio
${{\cal R}(v)}$ will not vary with $v$. But, in a 
scenario where the multi-phase structure resembles that of the ISM,
the Si II velocity profiles would be wider than {\\}  
the {\ciis} profile, since
$\sigma_{WNM}$$\approx$2$\sigma_{CNM}$ in 
the ISM (Kulkarni \& Heiles 1987).
In principle this is a signature of the WNM.
But differences between the {\em observed} profiles will be diluted
by the finite resolution of HIRES (FWHM$\approx$7$-$8 {\kms}),
which is comparable to the widths of most components (i.e., they
are unresolved), and Poisson noise. A detailed evaluation
of these effects will be discussed elsewhere. 

Figure 11c compares {\ciis} and resonance-line profiles for the DLA
toward Q2231$-$00. Though the profiles are noisier in this case,
some effects are clear. First, the ratio {\Rratio} does not exhibit
significant variations at $v$ $<$ $-$60 {\kms}.
On the other hand, there is evidence for variations of {\Rratio}
in the component centered at $v$ = 0 {\\} {\kms}. At $v$ $\approx$
$+$ 10 {\kms}, {\Rratio} is lower than $<$${\cal R}$$>$, the mean {\Rratio}
integrated over the entire profile (see panel 4 of Fig. 12c).
This is consistent with WNM gas at $v$ $\approx$
$+$ 10 {\kms}. At the same time, there is tentative evidence for
enhanced {\ciis} absorption between  $v$ $\approx$ $-$ 30 {\kms}
and $-$ 10 {\kms}
where {\ciis} 1335.71 exhibits stronger absorption 
than either Si II 1808
or Ni II 1741. While
weak
{\ciis} 1335.66 absorption at $v$ $\approx$ 0 {\kms}
may cause
the excess {\ciis} 1335.71 absorption at $v$ $\ge$ $-$ 10 {\kms}
it cannot explain the excess {\ciis} 1335.71 absorption at $v$ $<$ $-$
10 {\kms}.

Excess {\ciis} absorption
can be due to increased heating of the CNM. In the
case of absorption by CNM gas alone we have
\begin{equation}
{{\cal R}(v)}{\equiv}{{\tau_{v}}({\rm C II}^{*}) \over {\tau_{v}}({\rm SiII})} = {\Biggl[}{{f_{{\rm C II}^{*}}{\lambda_{{\rm C II}^{*}}} \over {f_{{\rm SiII}}{\lambda_{{\rm SiII}}}}}{\times}{1 \over {h{{\nu}_{ul}}A_{ul}}}}{\Biggr]}{\Biggl[}{{10^{-24}{\kappa}{\epsilon}G_{0}} \over ({\rm Si/H})}{\Biggr]}
\cmma
\label{eq:RvCNM}
\end{equation}

\noindent where we combined the definition of 
{\gamdnr} (equation 1 in Paper I) and equation
({\ref{eq:lcCNM}}). Detectable variations
in {\Rratio} in CNM gas are most likely caused by variations in {$\epsilon$}
and {$G_{0}$} rather than in absolute element abundances. This is because
\begin{equation}
{{\kappa} \over ({\rm Si/H})}={{\Bigl(}10^{[{\rm Fe/Si}]_{int}}-10^{[{\rm Fe/Si}]_{gas}}{\Bigr)} \over ({\rm Si/H})_{\odot}}
\cmma
\label{eq:kapovZSi}
\end{equation}

\noindent and the recent analysis by Prochaska (2002) showed evidence
for remarkable uniformity in the {\em relative} abundances of DLAs 
(with the exception of the DLA toward Q0347$-$38). 
As a result, 
\begin{equation}
{\delta}_{{\cal R}}(v)={{{\epsilon}(v)G_{0}(v)-<{\epsilon}G_{0}>} \over <{\epsilon}G_{0}>}
\perd
\end{equation}

\noindent where ${{\delta}_{{\cal R}}}(v)$=[{\Rratio}-$<${$\cal R$}$>$]/$<{\cal R}>$,
and $\epsilon (v)$ and $G_{0}(v)$ are the grain photoelectric heating efficiency and
FUV mean intensity at velocity $v$.
Because $\epsilon$ is a function of $G_{0}{\sqrt T}/n_{e}$ (Bakes \& Tielens 1994;
Weingartner \& Draine 2001a),
${{\delta}_{{\cal R}}}(v)$
will vary with $v$ if any one of
these parameters changes with $v$. If $G_{0}{\sqrt T}/n_{e}$
$<<$ 5$\times$10$^{3}$ K$^{1/2}$cm$^{3}$, the grains are mainly neutral and
$\epsilon$ is insensitive to variation in $G_{0}$. Therefore, in this
CNM limit,
${{\delta}_{{\cal R}}}(v)$
$\approx$ ${\Delta G_{0}}/G_{0}$. The excess
{\ciis} absorption in the Q2231$-$00 DLA can then result from
the incidence of a larger-than-average radiation 
intensity, $G_{0}$, 
on the CNM gas in the velocity range $v$ = $-$30 to $-$10 {\kms}.
Spatial variations in radiation intensity along the line
of sight could result from (a) nearer than average displacement of OB stars
from this gas, (b) differing CNM cloud  distances from a centrally-located
bulge source, or (c) in the case of CDM galaxy
formation scenarios, passage of the line-of-sight through separate protogalactic
clumps with independent SFRs (Haehnelt {\etal} 1998).
The CDM scenario should produce standard deviations,
$\sigma_{{{\delta}_{{\cal R}}}}$ $\approx$ 1,
because the SFRs
would be uncorrelated at the different
clump locations. But in the bulge scenario, we find
$\Delta G_{0}/G_{0}$ $\sim$ $h/b$ where $h$ is the thickness of the
surrounding disk and $b$ is the
sightline impact parameter. This would lead to 
$\sigma_{{{\delta}_{{\cal R}}}}$ $<$ 1,
for typical values of
$b$. To discriminate between these hypotheses, 
larger samples of DLAs with measured {\Rratio} are required.

\section{DO H IGH-$z$ DLAs CONTAIN CNM GAS ?}

In this section we discuss possible objections to the 
presence of CNM gas in DLAs. Recall that 
if {\ciis} 1335.7 absorption arises 
in WNM gas, the inferred FUV radiation contributes more
background radiation than observed. On the other hand if it
arises in CNM gas, the background radiation is consistent
with observations.
Nonetheless, the following
arguments 
have been made against the presence of CNM gas
in high-$z$ DLAs.

\subsection{C II/C I Ratios}

The first of these has to do with the
large values observed for the ratio, $N$(C II)/$N$(C I) 
(hereafter C II/C I). 
Liszt (2002) constructed two-phase models similar to ours

\begin{figure}[ht]
\includegraphics[height=4.2in, width=3.8in]{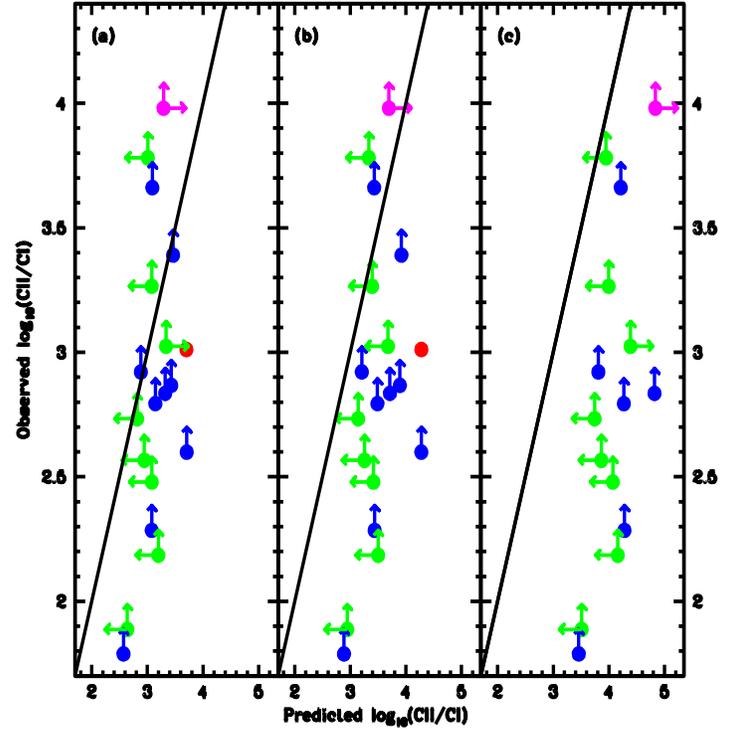}
\caption{Plot of observed versus predicted C II/C I ratios for
19 DLAs from Table 1. With the exception of the filled
red circle, which is a positive detection,
the {\em observed} C II/C I ratios are lower limits.
The filled blue circles correspond to DLAs
for which we have positive detections 
of {\lclos}. 
Filled green circles correspond to DLAs with limits
on {\lclos} and filled magenta circle corresponds
to DLA with lower limit on {\lclos} (see text).
(a) Results for model with log$_{10}$$N_{w}$=19.3 cm$^{-2}$
and standard ratio of $\zeta_{CR}$/{\ps} (equation 9 in 
Paper I). (b) Same as (a)
except log$_{10}$$N_{w}$=20 cm$^{-2}$. (c) Same as (b) 
except $\zeta_{CR}$/{\ps} equals 0.3 times the standard
ratio.}
\label{c2starratio}
\end{figure}

\noindent and computed C II/C I ratios for 5 DLAs for which observed
ratios were available. While the ratios he predicted
for the CNM were lower than observed,  
those predicted for the WNM were consistent with observed
values. He concluded the observed C II/C I ratios permit no more
than a few percent of the DLA gas to be in the CNM phase.

We repeated these calculations with the CNM model with
``Gal'' dust and minimal depletion.
In Figure 12 we compare the results with C II/C I
ratios deduced for 19 DLAs in our sample. With the exception
of the filled red circle depicting a positive detection,
the {\em observed} C II/C I ratios are lower limits,
while the {\em predicted} C II/C I ratios are definite
numbers if {\lclos} is detected (filled blue circles), and
upper or lower limits if limits are placed on {\lclos}
(filled green circles; see caption to Figure 12).
Figure 12a shows results for a model incorporating many
of the assumptions made by Liszt (2002). In particular,
we assume the incident soft X-ray radiation 
is attenuated by gas with H I column density
log$_{10}$$N_{w}$ = 19.3 cm$^{-2}$. We are in general
agreement with Liszt (2002) that this CNM model cannot
account for the large values observed for C II/C I. Although
the model predictions should lie to the right of the diagonal line
representing equality between observed and predicted ratios,
one third of the sample is on or to the left of the line. This
disagreement takes on significance when it is realized
that all the discrepant points are lower limits
on the observed C II/C I ratio and hence the actual
ratios are displaced even further from agreement with
model predictions.

However, the predicted C II/C I ratios are sensitive to
changes in model input parameters. This is evident in
Figure 12b showing results for the same model as
in Figure 12a except N$_{w}$ is increased
to 10$^{20}$ cm$^{-2}$; i.e., to the attenuating column density,
which is more appropriate for DLAs and which is used
in all our calculations (see
discussion in Paper I). Comparison with Figure 12a
reveals an increase in predicted C II/C I ratios by
$\sim$ 0.4 dex with a consequent improvement between
theory and observation. The reasons for the increase
in C II/C I
are as follows: The increase in $N_{w}$ results in a reduction
in X-ray intensity, which in turn causes a decrease in
the heating rate. This decreases the gas pressure,
which brings about a decrease in the CNM density, $n_{CNM}$
as illustrated in Figure 5 in Paper I.
For a given {\lclos} a decrease in $n_{CNM}$ causes an
increase in {\ps} (hence $G_{0}$), a decrease in $n_{e}$,
and a rise in $T$. All three factors conspire to increase
the C II/C I ratio, since it is proportional to
$G_{0}/(n_{e}{\alpha_{{\rm C I}}}(T))$,  where ${\alpha_{{\rm C I}}}(T)$,
the case A recombination coefficient to  C I, decreases
with increasing $T$.
For many DLAs even better agreement is achieved if
we retain log$_{10}$$N_{w}$ = 20 cm$^{-2}$ and reduce
the ratio of cosmic ray ionization rate to SFR
per unit area, $\zeta_{CR}$/{\ps}, below the ratio given in 
equation (9) in Paper I.
At these low X-ray intensities, cosmic rays still dominate the ionization
rate, and as a result $n_{CNM}$ is reduced even
further.
Figure 12c shows results for
$\zeta_{CR}$/{\ps} equaling 0.3 times the
ratio in equation (9) in Paper I. 
In this case most of the predicted C II/C I ratios
are in better agreement with observations
than before.
However, in the most metal rich
DLAs the additional decrease in pressure
accompanying the reduction in $\zeta_{CR}$/{\ps}
results in the disappearance of the pressure
maxima and minima essential for two-phase
equilibria. This is why four of the DLAs in
Figures 12a and 12b are missing from Figure 12c.
Clearly, more realistic models might include
a range in $\zeta_{CR}$/{\ps} ratios rather
than assigning the same value to each DLA.


The point of this exercise is to show that the 
C II/C I ratio depends sensitively on the X-ray and
cosmic ray ionization rates, both of which are
uncertain. For these reasons, 
it is premature to use the C II/C I ratio
to rule out CNM gas in DLAs. On the other
hand, the observed C II/C I ratios are useful
for placing upper limits on the 
$\zeta_{CR}$/{\ps} ratio. We find that
$\zeta_{CR}$/{\ps} in our model CNM gas cannot be more than 2 times
the value in equation (9) in Paper I.
Otherwise more than one third of the
points in the Figure 12 would lie above the diagonal line.

\subsection{Equilibrium Gas Pressures in DLAs}

The second argument against CNM gas in high-$z$ DLAs was made by
Norman {\&} Spaans (1997). They concluded that
high-$z$ DLAs instead consisted of pure WNM gas with
pressures $P$ $<$ $P_{min}$. Computing two-phase equilibria
in the context of CDM models for galaxy formation, they found
that at $z$ $>$ 1.5 the gas equilibrium pressure, $P_{eq}$, 

\begin{figure}[h]
\centering
\scalebox{0.33}[0.35]{\rotatebox{-90}{\includegraphics{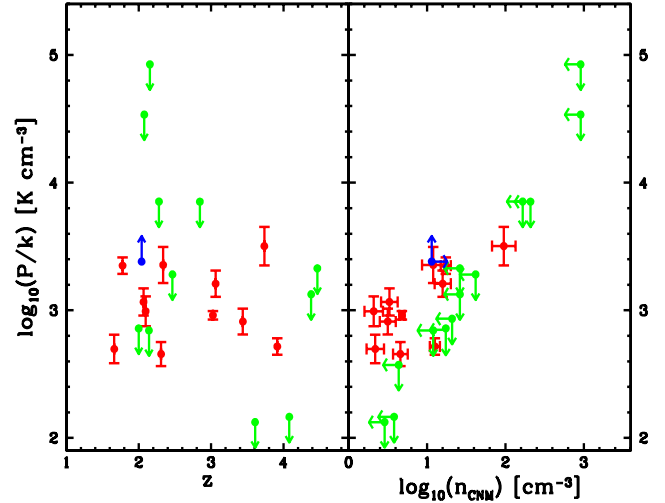}}}
\caption[]{Plot showing equilibrium pressure, 
(i.e., $(P_{min}P_{max})^{1/2}$), versus $z$ and $n_{CNM}$.
Pressures computed for standard CNM model with ``Gal'' dust
and minimal depletion. (a) Plot of $P/k$ versus $z$. Two
upper limits at log$_{10}(P/k)$ $<$ 2.2 K cm$^{-3}$ correspond
to DLAs along sightlines toward Q1108$-$07 and Q2237$-$06, which probably
encounter only WNM gas. In that case the pressures
would be higher than shown. (b) Plot of $P/k$ versus
$n_{CNM}$.}
\label{rhovszSMC}
\end{figure}

\noindent exceeded
the hydrostatic pressure, $P_{hydro}$, in the midplanes
of protogalactic disks embedded in dark-matter halos. By
contrast $P_{eq}$ was predicted to be less than
$P_{hydro}$ at $z$ $<$
1.5. They concluded that the hydrostatic pressure available
in model protogalaxies is insufficient to generate
two-phase media at $z$ $>$ 1.5. The ratio $P_{hydro}/P_{eq}$
decreases with increasing redshift 
because
$P_{eq}$ increases due to the sharp decline of metallicity with
redshift predicted by Norman \& Spaans (1997; Spaans 2002). 

We tested these predictions using measurements of {\lclos}
to infer $P_{eq}$. The results shown in Figure 13 were obtained
with the CNM, ``Gal'' dust, minimal depletion model discussed
above. In Figure 13a we plot $P_{eq}$ versus $z$.
The average of the pressures corresponding to the positive
detections is $P_{eq}$=(1.33$\pm$0.81)$\times$10$^{3}$ K cm$^{-3}$.
This is much lower than the gas pressures predicted by
Norman \& Spaans (1997), which exceed $\sim$ 10$^{5}$ K cm$^{-3}$ at
$z$ $>$ 1.5 (see their Figure 3), but in agreement
with $P_{hydro}$ predicted for typical
CDM models.
According to the CDM models of Mo {\etal} (1998), 2/3 of DLAs
detected in absorption at $z$ $\approx$ 2.5 should arise
in halos with circular velocities, $V_{c}$ $>$ 100 {\kms}.
Using the Norman \& Spaans (1997) formalism we find 
the corresponding hydrostatic pressures to exceed
2$\times$10$^{3}$ K cm$^{-3}$. Given the uncertainties,
we conclude that the hydrostatic 
pressures available in CDM models are sufficient to
generate the type of two-phase equilibria we infer
for DLAs. Coincidentally, these pressures could also
arise in high-$z$ DLAs if they
resemble nearby massive galaxies such as the Milky Way where the
gas 
pressures at the solar circle are 
$\approx$ 3$\times$10$^{3}$ K cm$^{-3}$
(Wolfire {\etal} 2002). 
The analysis of Mo {\etal} (1998) and Norman \& Spaans (1997)
shows
$P_{hydro}$ $\propto$ $[V_{c}$$H(z)]^{2}$, where 
the Hubble parameter
$H(z)$ is an increasing function of redshift.
Therefore low-mass objects at high redshift
can in principle generate as much hydrostatic pressure as
high-mass objects with low redshifts.
As a result,
pressure does not discriminate between
DLA models based on
CDM from the null hypothesis in which DLAs are
the unevolved disks of current
normal galaxies (e.g. PW97). Nor does
the absence 
of redshift evolution evident in Figure 13a. This is an obvious prediction
of the null hypothesis. In the case of CDM we note that
the non-linear mass scale,  $M_{c}$ $\propto$ (1+z)$^{-(6/(n+3))}$,
where $n$ is the power-spectrum index (Padmanabhan 1993), 
and the virial velocity, $V_{c}$ $\propto$ $[H(z)M_{c}(z)]^{1/3}$,
where the Hubble parameter, $H(z)$ $\propto$ $(1+z)^{3/2}$ for
the Einstein-deSitter cosmology.
Since $n$ = $-$ 2 at galactic scales, 
$P_{hydro}$ is independent of redshift. As a result,
the redshift dependence 
of $P_{hydro}$ is negligible in both models.

In Figure 13b we plot $P_{eq}$ versus CNM density,
$n_{CNM}$. The densities corresponding to positive detections
(red data points) range between 2 and 100 cm$^{-3}$
with an average $<n_{CNM}>$ = 16 cm$^{-3}$. These resemble
densities inferred for the CNM clouds in the Galaxy
(W95). 
Because the H I column density of a typical DLA velocity 
component is approximately  1$\times$10$^{20}$ cm$^{-2}$
(PW97), the volume
densities imply physical dimensions on the order of
a few pc. This conclusion differs
from arguments that DLA clouds causing 21 cm absorption
uniformly cover the cores of
compact radio sources, which typically have linear sizes
of $\sim$ 100 to 400 pc 
(Briggs \& Wolfe 1983), and
has important implications for interpreting measurements
of 21 cm absorption in DLAs (see $\S$ 5.3).
The linear correlation evident in Figure 13b 
is tentative, because the Kendall $\tau$ test for
positive detections indicates
$\tau$ = 0.45 and the probability for the null hypothesis 
of no correlation, $p_{Kendall}$ = 0.052.
We find the average temperature $<T>$=190$\pm$130 K
for this model. 
The temperatures are higher than diffuse clouds in
the ISM due to the combination	
of low densities and low metallicities.

{\subsection{High Spin Temperatures at Large Redshifts}}

The third argument against CNM gas at high redshifts stems from the high
spin temperatures, $T_{s}$, deduced from 21 cm absorption observations of
DLAs. Whereas spin temperatures in nearby spirals are less than 300 K 
(Dickey \& Lockman 1990), the spin temperatures in most
DLAs exceed 500 K (Chengalur \& Kanekar 2001).
The discrepancy is greatest
at $z$ $>$ 3 where several DLAs
exhibit
$T_{s}$ $>$ 2000 K. 
Because the kinetic temperature, $T$, equals
the spin temperature in most scenarios, the high values of $T_{s}$ have
been interpreted as indicators of gas in the 
WNM rather than CNM phase (Carilli {\etal} 1996; Kanekar \& Chengalur
2001; Chengalur \& Kanekar 2002). This poses an interesting dilemma,
as {\ciis} absorption, which must arise in the
CNM in most DLAs, is detected in two DLAs with high
inferred spin temperatures.

How do we reconcile the spin temperature limit, 
$T_{s}$ $>$ 4000 K (Ellison {\etal} 2001; Kanekar \& Chengalur 2001),
with the  $\approx$ 100 K temperature
predicted for the CNM in the $z$ = 3.387 DLA toward Q0201$+$11?
To answer this question we consider the sizes of the CNM
clouds relative to the background radio source in Q0201$+$11.
Following the discussion in $\S$ 5.2 we find that the linear
dimensions of CNM clouds are less than 10 pc,
since
$n_{CNM}$ $\approx$ 6 cm$^{-3}$ and we have assumed
half the total H I column density of 1.8$\times10^{21}$ cm$^{-2}$ 
to be CNM gas equally distributed among 5 or more velocity
components. To estimate the linear dimension subtended by
the radio source at the DLA we note that VLBI observations
at $\nu$ = 1.6 GHz show Q0201$+$11 to subtend a solid
angle less than $\omega_{1.6}$=2.5$\times$5.0 mas$^{2}$ 
(Hodges {\etal} 1984). The solid angle of the source
must be larger than $\omega_{1.6}$ at $\nu$ = 323.7 MHz,
the frequency of redshifted 21 cm absorption, otherwise
the brightness temperature of this radiation in the rest frame of the
$z$ = 3.61 QSO, $T_{b}$(1492 MHz)=1.7$\times$10$^{12}$K. This
exceeds the 10$^{12}$K Compton limit restricting the brightness
temperature of sources such as Q0201$+$11, which belongs to the class of peak
spectrum objects {\em not} exhibiting relativistic beaming
(Phillips \& Mutel 1982). To be consistent with the 
Compton limit the
source must subtend an  effective linear diameter {\em exceeding}
40 pc at the DLA (where we assume {\omgm}=0.3, {\omgv}=0.7, and $h$=0.7). 
As a result the typical CNM cloud in this DLA covers
a small fraction of the background radio source. This
contrasts with the conclusions of deBruyn {\etal} (1996)
who did not consider the Compton limit.

It is now possible to understand the discrepancy between
$T_{s}$ and the temperature of the CNM. In the optically thin
limit 
the apparent
21 cm optical depth $\tau$(21)=$f_{CNM}${$\times$} \\
${\tau}_{CNM}$(21)
where $f_{CNM}$ is the area covering factor of CNM gas,
${\tau}_{CNM}(21)$ is the 21 cm optical depth of a CNM cloud,
${\tau}(21)$$\equiv$${\ln}$[$S_{c}/S_{v}$], and 
$S_{v}$ and $S_{c}$ are the 373.3 MHz flux densities at Doppler velocity $v$
and in the continuum respectively. Chengalur \& Kanekar (2000)
placed a 1-$\sigma$ upper limit of 0.011 on ${\tau}(21)$. We 
shall be more conservative and assume a 95 $\%$ confidence
upper limit of ${\tau}(21)$ $<$ 0.022, which reduces the lower
limit on $T_{s}$ to 2000 K. Because
half the sample DLAs show positive evidence for 
{\ciis} absorption, while in 3 out of 25 DLAs there is 
probable evidence for WNM gas (see Paper I), 
we find 0.5 $<$ $f_{CNM}$ $<$ 0.88.
We shall assume $f_{CNM}$ = 0.67 as this is in accord with
the observed relative
occurrence of multiple {\ciis} velocity components in DLAs. 
As a result we find
${\tau}_{CNM}(21)$ $<$ 0.033.
To compute $N_{CNM}$(H I),
the H I column density of the CNM cloud, we must estimate $\sigma$, the
Gaussian velocity dispersion of the gas, since
$N_{CNM}({\rm H I})$=
1.82$\times$10$^{18}$$\times$${\tau}_{CNM}(21)$$T_{CNM}$${{\sqrt {2{\pi}}}}{\sigma}$.
Assuming $\sigma$ = 9 {\kms}, the velocity dispersion corresponding
to the velocity width to which the upper limits on $\tau$(21) apply, 
and $T_{CNM}$ {$\approx$} 100 K,
we find $N_{CNM}({\rm H I})$ $<$ 1.4$\times$10$^{20}$ cm$^{-2}$. 
This is consistent with a total CNM H I column 
toward the optical continuum  of
9{$\times$}10$^{20}$ cm$^{-2}$ distributed among 5 or more
components, and a typical CNM HI column of 
$<$ 1.4$\times$10$^{20}$ cm$^{-2}$ covering the radio source.
In other words, the {\em inferred} spin temperature for the DLA
toward Q0201$+$11 is high not because the temperature
of the absorbing gas is high, but rather because the H I column
densities of the foreground CNM clouds are low.
In fact there is evidence for a decrease
in {\NH} with increasing redshift. Peroux {\etal} (2002)
find the mean {\\} H I column density of DLAs, $<N({\rm H I})>$
= 3.9$\times$10$^{20}$ cm$^{-2}$ at $z$ $>$ 3.5, while
$<N({\rm H I})>$ = 8.2$\times$10$^{20}$ cm$^{-2}$  at
$z$ = 2.4$-$3.5. Because these correspond to the total 
{\NH} rather than the CNM portion of the gas, it is plausible to
assume the mean H I column densities of CNM gas to be
2$\times$10$^{20}$ cm$^{-2}$ at the DLA redshift, $z$ = 3.38.
Therefore, it is not improbable for {\NH}$_{CNM}$ to equal
(1$-$2)$\times$10$^{20}$ cm$^{-2}$ in this DLA. If this interpretation
is correct, the high H I column density ({\NH}
= 1.8$\times$10$^{21}$ cm$^{-2}$) and {\\} 
{\ciis} detected  in 
absorption toward 
Q0201$+$11 indicate that a larger than average 
number of CNM clouds are lined up
toward the {\em optical} continuum source of this QSO.
This is supported by the excessive number of
low-ion clouds spread across a velocity interval
of $\sim$ 270 {\kms} (Ellison {\etal} 2001);
by comparison the median velocity interval of DLAs is $\approx$ 100 {\kms} (PW97). Note, Chengalur \& Kanekar (2000) would not have
detected this configuration, since it would have an apparent
optical depth, $\tau$(21) = (10pc/40pc)$^{2}$
{$\times$}5{$\times$}${\tau}_{CNM}(21)$ $<$ 0.01, which is
less than the 2-$\sigma$ upper limit of 0.022.

Ellison {\etal} (2001) carried out an imaging study 
of the field surrounding Q0201$+$11 that provides a test
of the CNM hypothesis. 
If {\ciis} absorption in this DLA arises in WNM gas, the FUV
intensities (at $\lambda$ = 1500 {\AA}), $J_{\nu}$,
would be higher than indicated by our CNM model. 
We find  
({\jnu})$_{WNM}$=3.8$\times$10$^{-18}$ {\junit}, while
({\jnu})$_{CNM}$=4.3$\times$10$^{-19}$ {\junit}  for the
case of ``Gal'' dust and maximal
depletion. 
Based on photometric redshifts, the leading candidate for the galaxy
responsible for DLA absorption is object G2, the high surface-brightness
region of an $R$ $\approx$  25.3  galaxy
separated by $\Delta \theta$ = 2.9 arcsec from the QSO.
If the FUV radiation
emitted by G2 heats WNM gas responsible for {\ciis} absorption, according
to the bulge
hypothesis G2 would detected in the $R$ band
with flux density, $S_{{\nu}_{0}}$=4{$\pi$}{\jnu}(${\Delta \theta})^{2}$/(1$+$$z_{DLA}$)$^{3}$,
where ${\nu}=(1+z_{DLA}){\nu}_{0}$. In this case the predicted AB 
magnitude, $R$ = 21.3. If instead this WNM gas is heated by a uniform disk
of sources centered on G2 and extending across the QSO sightline,
the surface brightness of the disk
would be ${\mu}_{R}$=26.4 mag arcsec$^{-2}$. Both WNM scenarios are
ruled out by the Ellison {\etal} (2001) images. In the bulge scenario,
G2 would be 4 magnitudes brighter than observed,
while in the 
uniform disk scenario the 
surface brightness  of the disk would be detectable
at the 4-$\sigma$ level
in regions on the side of G2 away from the QSO.
On the other hand if G2 heats CNM gas responsible for {\\}
{\ciis} absorption,
then $R$ = 23.6 in the bulge scenario, while
${\mu}_{R}$=28.7 mag arcsec$^{-2}$ 
in
the uniform disk scenario. 
Because of the large uncertainties in the $R$ magnitude
of G2, 
the bulge scenario is only
marginally consistent with the data, while the uniform
disk scenario is definitely consistent with the data. Therefore, the
gas producing {\\} {\ciis} absorption must be a CNM if it is heated
by FUV radiation emitted
by sources associated with G2. Of course,
G2 may be incorrectly
identified and the galaxy associated with the DLA could be located
within the PSF of the QSO (Ellison {\etal} 2001), 
in which case {\ciis} absorption could arise in either phase. 
This emphasizes the importance of obtaining a spectroscopic
redshift for G2.

More recently, Kanekar \& Chengalur (2003) deduced a 95 $\%$ confidence
lower limit of $T_{s}$ $>$ 1.4{$\times$}10$^{4}$ K for the
$z$=3.062 DLA toward Q0336$-$01. By contrast we infer an equilibrium
temperature of $\approx$ 100 K from the presence of {\ciis} absorption
in this DLA (see Table 1 in Paper I). In principle, we can explain the high
inferred value of $T_{s}$
with low values of {\nh}$_{CNM}$ in this object
also. However, much lower CNM filling factors are required
due to the higher value of $T_{s}$. We think it more likely
that the high $T_{s}$ inferred for this DLA is a byproduct of a
complex radio-source
structure. Though this source is observed to be compact at 
5 GHz, it is likely
to be extended at the much lower frequency predicted for redshifted 21 cm
absorption. In fact, it is possible that the radio emission
is concentrated in two equally bright components, each of which
is symmetrically displaced more than 10 pc
from the optical continuum (see Phillips \& Mutel 1982).
In that case the gas detected in optical absorption need not
intercept radio-frequency radiation along the line of sight. In any
case  it is difficult to understand how
the lower limit on $T_{s}$ can correspond to the temperature of 
neutral gas in any DLA,
since 50 $\%$ of
H is collisionally ionized at 
1.5$\times$10$^{4}$ K.

\section{SUMMARY and CONCLUDING REMARKS}
This paper expands on a new technique for measuring SFRs in DLAs
developed in Paper I. Namely, in Paper I we showed how detections of {\ciis}
1335.7 absorption could be used to infer the SFR per unit physical
area, {\ps}, in DLAs. We showed that a two-phase neutral medium,
in which CNM gas is in pressure equilibrium with a WNM gas, is
a natural byproduct of thermal equilibrium. We also found
that {\ciis} absorption lines detected in DLAs could arise in
either phase, but that significantly higher values of {\ps}
were required if this absorption arose in the WNM. In this
paper we use cosmological constraints to show that while
the line-of-sight likely encounters both phases, {\ciis}
absorption arises mainly in the CNM. Our specific
conclusions are as follows:

(1) We compute the SFR per unit comoving volume
in DLAs, {\rhodotz}, by
combining the mean SFR per unit area, {\psavz},  with the
observed number of DLAs per unit absorption distance.  We obtain 
statistically significant results for two redshift bins centered
at $z$ = 2.15 and $z$ = 3.70; these are the first
quantitative measurements of {\rhodot} at $z$ $\approx$ 2. The results
show (a) {\rhodotz} for the WNM model is at least 10 times higher
than for the CNM model, (b) {\rhodotz} for the CNM model is
in approximate agreement with independent determinations from
luminosities measured for flux-limited samples of galaxies, and
(c) no evidence for evolution in the redshift interval, $z$ = [1.6, 4.5].
We also compute the bolometric background intensities, $I_{EBL}$,
generated by the SFR histories, {\rhodotz}. In every case the
WNM models predict $I_{EBL}$ above the observed 95 $\%$ confidence
upper limits. By contrast, the $I_{EBL}$ predicted for the CNM
models are consistent with this limit. As a result, models in which
{\ciis} absorption arises in WNM gas are ruled out. Finally,
we develop a ``consensus'' model, which 
accounts for the systematic errors arising from
various model uncertainties. 
We also find
{\psavz} for DLAs appears to decrease significantly with
decreasing redshift
at $z$ $<$ 1.6 if DLAs evolve into ordinary galaxies.
To compute {\psavz} we average {\ps} over $R_{HI}$ rather than
the de'Vaucouleurs radius.

(2) We consider several consequences of our work. First, we evaluate the
mass of stars
and metals produced by the star formation histories we derived. 
Though the mass in stars produced by $z$ = 0 is consistent with
the masses of current stellar populations in current galaxies, the
mass of metals produced by $z$ = 2.5 is significantly higher than is observed for
DLAs at that redshift. Various solutions to this dilemma come
to mind, including ejection of metal-enriched gas
from DLA, making DLAs a transitory phase
of galaxy evolution. But this results in an IGM metallicity, [M/H] = $-$1.2,
which is a factor of 100 larger than observed
in the {\lya} forest. Rather we favor a scenario
in which star formation and metal production occur in a centrally-located
bulge region displaced from the DLA gas.  Second, we evaluate the 
bulge hypothesis and find that to within 10 $\%$ accuracy the
predicted {\rhodotz} agrees
with that predicted by the uniform-disk model used to derive
the results discussed above. From 
the empirical upper limit on $I_{EBL}$, we find that the fraction of 
FUV radiation attenuated by optically thick dust in the bulge cannot
exceed 0.7. Third, we search for evidence of connections 
between stars and gas. We find no evidence for a Kennicutt (1998) correlation
between {\ps} and {\NH}. This is naturally explained by the bulge
hypothesis in which SFRs in the bulge region are unrelated to
the H I column density in regions giving rise to DLA absorption.
The lack of a Kennicutt relation is also consistent with the
model in which star formation occurs in the same region creating 
DLA absorption. This is because {\ps} is the SFR averaged over
the entire DLA 
(5 kpc or more in most models), while {\NH} is sampled over a transverse dimension
corresponding to the linear diameter of the QSO, i.e., $\sim$ 1 pc.
On the other hand when
these quantities are averaged over the DLA sample, they are
consistent with the Kennicutt
relation, indicating that it may be present in a statistical sense.  
We also look for correlations between {\ps} and metallicity, and
{\ps} and kinematic velocity width. Marginal evidence for correlations
were found in the cases of metallicity and low-ion
velocity width.  Confirmation would indicate that both low-ion line
width and metallicity are global parameters, which are determined
by quantities such as dark-matter mass. 
The reasons for  null 
correlations in the cases of {\ps} and high-ion
velocity width are the same as for the Kennicutt relation.

(3) We discussed tests of the ideas presented here. 
First, we present 
statistically significant evidence for
a correlation between {\lclos} and {\kapnr}. This is consistent with
grain photoelectric heating by DLAs with a limited range of SFRs,
since {\lclos} $\propto$ {\kapnr}{$\epsilon$}{\ps}. We show that this
correlation is not naturally explained by alternative heating mechanisms
such as cosmic-ray ionization and X-ray photoionization. 
Second, we consider the possibility that the $^{2}P_{3/2}$ and $^{2}P_{1/2}$
fine-structure states in C$^{+}$ are populated by CMB radiation rather
than collisions. In that case {\lclos} would not equal the cooling rate as
we have assumed, but would instead reflect the local temperature of the
CMB. We test this hypothesis by comparing the observed 
{\lclos} with 
({\lclos})$_{CMB}$, the
spontaneous energy emission rate per H atom when CMB radiation alone
populates the fine-structure states. We conclude that the CMB alone
cannot explain the observed level of {\ciis} excitation in 28 out of 30
DLAs. This is because the observed {\lclos} do not show
the sharp increase with redshift predicted for ({\lclos})$_{CMB}$
and because in all 28 cases, {\lclos} $>>$ ({\lclos})$_{CMB}$.
The two exceptions are DLAs with upper limits on {\lclos}, and in which
({\lclos})$_{CMB}$ $>$ 0.8{\lclos}. The implied cooling rates
are very low, and may indicate
passage of the QSO sightlines through a WNM subjected to
SFRs less than or equal to the mean {\psav}  deduced for most of our sample. 
In the third test, we use
the ratio of  {\ciis} to resonance-line optical depths,
${\cal R}(v)$ = $\tau_{v}$({\ciis})/$\tau_{v}$(Si II), to probe the
multi-phase structure of the gas. Specifically,
${\cal R}(v)$ in the WNM should be ({\lclos})$_{WNM}$/({\lclos})$_{CNM}$
times ${\cal R}(v)$ in the CNM. Because ({\lclos})$_{WNM}$/({\lclos})$_{CNM}$
$\approx$ 0.08, ${\cal R}(v)$ should decrease significantly at velocities
corresponding to the WNM. Such variations may have been detected in 
one DLA. We also show how the WNM is easy to hide, and we discuss
possible evidence for an increase in ${\cal R}(v)$ at velocities
where {\ps} in the CNM increases. 

(4) We discuss possible objections to our picture of DLAs.
The most persistent of these is the proposal that DLAs do not
contain CNM gas, but rather are comprised only of WNM gas.
The first argument against the CNM is related to the
large  C II/C I ratios detected in DLAs, since this
ratio is predicted to be lower in the CNM of the Milky Way ISM.
However, this conclusion is sensitive
to the ratio of X-ray and cosmic-ray heating rates to the 
SFR per unit area. In our models the X-ray heating rate is
considerably lower than in the ISM.
For our model parameters we are able to construct
reasonable models predicting C II/C I
observations. The next argument against the CNM is that 
the hydrostatic pressures available in low-mass galaxy 
progenitors predicted for CDM theories are significantly
lower than the values of $P_{min}$ predicted for two-phase
models. In that case, CDM models cannot generate sufficient pressure
to create a two-phase medium. We show this model is incorrect
because the high values of $P_{min}$ were based on an underestimate
of the DLA metallicities at large redshifts.
The third argument against the CNM stems from the high spin
temperatures deduced from the low 21 cm optical depth, $\tau(21)$
$<$ 0.02, and high {\NH} (= 2$\times$10$^{21}$ cm$^{-2}$) of
a high-redshift DLA. How does one reconcile the large inferred
spin temperature ($T_{s}$ $>$ 2000 K) with the low temperatures
($T$ $\sim$ 100 K) indicated by the detection of
{\ciis} absorption in this DLA? This answer is related
to the geometry  and H I column densities, {\NH}$_{CNM}$, of
the CNM clouds. Specifically, our models predict CNM clouds
with  sizes less than 10 pc, which is smaller than the
diameter of the background radio source. In that case 
we find {\NH}$_{CNM}$ $\sim$ 10$^{20}$ cm$^{-2}$ 
for typical CNM clouds covering the 
high-$z$ background radio source. Therefore,
the inferred spin temperature is high not because the
temperature of the absorbing gas is high, but rather because
{\NH}$_{CNM}$ is low. 
As a result, the large value of {\NH} toward the optical
continuum source implies
a larger than average number of CNM clouds along this line of sight,
which is observed.

While we have attempted to evaluate important
sources of systematic error, uncertainties still remain in our
analysis. 
For example, we cannot rule out the possibility that the
grain size distribution is radically different from the
MRN model used in our calculations. However, we think this is unlikely,
as the MRN distribution nicely describes grain properties in
local group galaxies, which are linked 
to high-$z$ DLAs by a similarity in the
comoving density of stars in the galaxies to the
comoving density of neutral gas in the DLAs. While the composition of the
dust is also uncertain, sensitive searches for the
2175 {\AA} feature will greatly
reduce this source of systematic error.
Secondly, though we have presented
evidence for a two-phase medium in our DLA sample, in particular
we described evidence for the presence of a WNM in 3 DLAs, the
thermal state of the gas may differ from the equilibria envisaged
in the models. Heiles (2001) has argued that warm clouds
in the ISM have temperatures of $\approx$ 2000 K rather than
the 8000 K predicted for the WNM. However, the
{\ciis} absorption we observe is unlikely to arise in such gas 
because at $T$ $>$ 1000 K the ratio of {\lclos} to the total cooling (i.e., heating)
rates is only slightly larger than predicted for the WNM.
As a result, SFRs similar to those predicted for the WNM would be required 
and would likely exceed
the observed upper limits on the bolometric background radiation
intensity.  There may be other caveats we have not discussed,  but they
are not obvious to us at this time.

Finally, we discuss
the similarity between {\rhodotz}
for DLAs and 
LBGs. Does it   
mean they are 
the same
objects?  Schaye (2001) suggested they were, and that DLAs were
created by the intersection between the line of sight to the QSO
and gas outflow from foreground LBGs. He argued that since LBGs brighter
than ${R}$ = 27 have comoving 
densities $n_{co}$=0.016$h^{3}$ Mpc$^{-3}$,
they could account for the incidence of DLAs provided their H I
cross-sectional radius $r$=19$h^{-1}$ kpc for a cosmology with
({\omgm},{\omgv})=(0.3,0.7). With our measurement of {\rhodot}
we can compute the SFR per DLA, {\ms}, because {\rhodot}={\ms}$n_{co}$.
From Table 1 we find {\ms}=40 {\smpy} for $h$=0.7. SFRs this
high are ruled out by results from H$\alpha$ imaging surveys
of DLAs (Bunker {\etal} 2001; Kulkarni {\etal} 2001). Furthermore,
in a recent search for a cross-correlation between DLAs and LBGs,
Adelberger {\etal} (2002) found tentative evidence that DLAs 
with {$R$} $<$ 25 were 
more weakly clustered than LBGs with
{$R$} $<$ 25. 
As a result, current evidence
suggests that DLAs and LBGs brighter than ${R}$=27
are drawn from separate parent populations. But more evidence
is needed before this can be accepted as a robust result. 
If this is 
confirmed, the similarity between the 
SFRs per unit comoving volume of LBGs and DLAs would be a coincidence.


\acknowledgements

The authors wish to extend special thanks to those of
Hawaiian ancestry on whose sacred mountain we are
privileged to be guests. Without their generous hospitality,
none of the observations presented here would have been possible.
We wish to thank 
Chris McKee for many valuable discussions about multi-phase media. We 
also thank
Bruce Draine, Eli Dwek, Mike Fall, Rob Kennicutt, Alexei Kritsuk, Jim Peebles,
Blair Savage, Marco Spaans,
Alexander Tielens, and Mark Wolfire for valuable comments.
We are grateful to A. Silva and S. Viegas for sending us their program, POPRATIO,
and to Len Cowie, 
Wal Sargent, and Tony Songaila for giving us data prior
to publication. 
Finally we thank
Simon White for remarks that stimulated this research.
A.M.W. was partially supported by NSF grant
AST 0071257.

\end{document}